**Convective Heat Transfer Optimization for Liquid Cooling Plates Driven by Field Synergy and Fractal Geometry**

Zixu Han, Peng Zhang*

Institute of Refrigeration and Cryogenics, Shanghai Jiao Tong University, Shanghai 200240, China

## Abstract

The rapid development of liquid-cooled data centers has imposed imperative demands on the performance of liquid cooling plate. The density-based topology optimization (TO) is an effective approach to resolving the growing thermal-hydraulic performance requirements of liquid cooling plate. However, existing TO methods can hardly optimize convective heat transfer directly which is the intrinsic heat transfer mechanism, due to the highly complex and evolving structural topologies, varying flow and temperature fields, making it extremely challenging to explicitly describe the heat transfer coefficient and heat transfer area during TO process. A convective heat transfer topology optimization (CTO) method is proposed in this study, where the iteratively evolving heat transfer coefficient is explicitly depicted by the field synergy theory in the thermal objective, and directly described by the velocity and temperature fields without relying on specific geometry. Combined with the explicit depiction of heat transfer area by the fractal geometry theory, a CTO framework is built for a direct optimization of convective heat transfer under both the laminar and turbulent flow conditions. The CTO tends to generate more hierarchical and directional structural topologies in optimization results, which is conducive to reducing low-velocity stagnation zones and improving flow direction in branched channels, achieving enhanced synergy and thermal-hydraulic performance in the optimized liquid cooling plates. Compared with the TO results without incorporation of field synergy theory, the CTO can reduce average temperature rise by 20% while improving the Nusselt number by 15% under laminar flow conditions, and reduce maximum temperature rise by 10.2% and pressure drop by 25% under turbulent flow conditions. The CTO framework enhances the synergy of optimization across various regions, effectively preventing the formation of optimized structures that only favor local heat dissipation while compromising

* Corresponding author: zhangp@sjtu.edu.cn

This article has been accepted for publication in Energy. The final published version is available at: https://doi.org/10.1016/j.energy.2026.142364

overall heat transfer, thereby proactively circumventing local optimal solutions.

## Nomenclature

| | | | |
|---|---|---|---|
| $A$ | specific area, $m^{-1}$ | $k$ | turbulent kinetic energy, $m^2/s^2$, thermal conductivity, W/(m·K) |
| $A_{eff}$ | effective heat transfer area, $m^2$ | $L$ | inlet width used as characteristic length for 2D topology optimization, m |
| $A_p$ | area of the design domain, heated area, $m^2$ | $Nu$ | Nusselt number |
| $c_A$ | constant in calculating specific area | $Nu^*$ | Nusselt number used in optimization |
| $c_h$ | constant in calculating heat transfer coefficient, $W\cdot s\cdot m^{-4}\cdot K^{-2}$ | $p$ | pressure, Pa |
| $C_{1\varepsilon}$ | constant in turbulence equation | $\Delta P$ | pressure drop, Pa |
| $C_{2\varepsilon}$ | constant in turbulence equation | $Pr$ | Prandtl number |
| $C_d$ | drag coefficient | $q$ | heat flux, $W/cm^2$ |
| $C_p$ | specific heat capacity, J/(kg·K) | $q_a$ | penalty factor |
| $C_{turb}$ | constant in turbulence generation term | $Q$ | volumetric heat source, $W/m^3$ |
| $D_f$ | local fractal dimension | $r$ | filter radius, m |
| $D_h$ | hydraulic diameter, m | $Re$ | Reynolds number |
| $Da$ | Darcy number | $Re^*$ | Reynolds number used in optimization |
| $f_p$ | flow resistance, $N/m^3$ | $s$ | fractal dimension related parameter |
| $g_\varepsilon$ | generation term of turbulent dissipation rate, $m^2/s^4$ | $T$ | temperature, K |
| $g_k$ | generation term of turbulent kinetic energy, $m^2/s^3$ | $V$ | velocity, velocity magnitude, m/s |
| $h_{avg}$ | average heat transfer coefficient, $W/(m^2\cdot K)$ | $w$ | weight parameter of objective function |
| $h^*$ | heat transfer coefficient used in optimization, $W/(m^2\cdot K)$ | | |

## Greek letters

| | | | |
|---|---|---|---|
| $\varepsilon$ | turbulent dissipation rate, $m^2/s^3$ | $\Omega$ | design domain |
| $\varphi$ | local objective function | $\rho$ | density, $kg/m^3$ |
| $\Psi$ | objective function | $\sigma_\varepsilon$ | constant in turbulence equation |
| $\gamma$ | design variable | $\sigma_k$ | constant in turbulence equation |
| $\mu$ | dynamic viscosity, Pa·s | $\theta$ | average synergy angle, ° |

**Acronyms**

| | | | |
|---|---|---|---|
| CTO | convective heat transfer topology optimization | PEC | performance evaluation criterion |
| FGTO | fractal geometry topology optimization | TO | topology optimization |

**Subscripts**

| | | | |
|---|---|---|---|
| $\beta$ | projection point | $p$ | projection |
| $f$ | fluid, hydraulic | $r$ | reference |
| $in$ | inlet | $s$ | solid |
| $l$ | laminar | $t$ | thermal, turbulent |
| $m$ | average | $top$ | interface between fluid and top heated plate |
| $max$ | maximum | $wall$ | average on the wall |
| $out$ | outlet | $0$ | initial value |

## 1. Introduction

With the rapid development of the artificial intelligence market, the surging demand for computing power has led to substantial energy consumption and carbon emissions from data centers. The data centers consume approximately 1.5% of global electricity, and the power consumption is projected to a rise by about 160% by 2030 [1, 2]. Since approximately 40% of the energy consumption of data center is used for cooling, efficient thermal management is critical for improving the power usage effectiveness of the data centers [3]. Furthermore, the increasing power density of electronic components like the CPUs and GPUs has led to continuously rising thermal load, further challenging thermal management technologies [4]. Conventional air-cooling system has become incapable of meeting the heat dissipation requirements as the heat flux increases, and are therefore gradually replaced by liquid cooling [5, 6]. Compared with other liquid cooling techniques, such as spray cooling or immersion cooling, indirect liquid cooling using liquid cooling plates is the most widely adopted due to its low cost, compact design, high reliability and adaptability [7-9].

The design of liquid cooling plates should consider both the heat dissipation capacity and pumping power consumption. Conventional optimization approaches, e.g., modifying size and shape of fins by simply varying geometric parameters, or directly adopting the biomimetic configurations [10], highly rely on experience and intuition and offer limited design flexibility. Consequently, it struggles to effectively balance thermal-hydraulic performances of liquid cooling plates to achieve an efficient optimization [11, 12].

Recently, the rapid advances in additive manufacturing have made it feasible to precisely manufacture liquid cooling plates with high structural complexity. Consequently, the topology optimization (TO) has become an effective approach for designing high-performance liquid cooling plates, due to its extreme design freedom and minimal susceptibility to manual intervention [13, 14]. Compared with other TO methods, such as the level-set-based TO, the density-based TO method requires lower computational costs, and it can restrict minimum channel dimensions more readily and ensure mesh-independency by incorporating techniques like filtering and projection [15, 16]. Therefore, most TO studies on liquid cooling plates design were conducted using the density-based TO methods.

Chen et al. [17] conducted single-objective TO on both the rectangular and serpentine channel cold plates with the target of maximizing heat dissipation rate, leading to temperature variance reductions of 19.5% and 41.8%, respectively. Xia et al. [18] investigated performance of bi-objective TO liquid cooling plates under various inlet and outlet structures, considering both the heat transfer and pumping power. It was found that the optimal overall performance of the optimized liquid cooling plates was achieved under the flared inlet-outlet configuration, with a 53.28% improvement in $Nu$ and a 40.89% reduction in $\Delta P$ compared to the aligned pin fins. Wu et al. [19] optimized the average temperature and pumping power of liquid cooling plates, while introducing a fitted correction function to account for the effects of non-uniform heat sources. Compared to the rectangular channel liquid cooling plate, the maximum temperature rise of the optimization results was reduced by 33.48% and the pressure drop was reduced by 61.96%. Liu et al. [20] systematically compared TO results under various aspect ratios of design domain, inlet-outlet numbers, inlet Reynolds numbers and thermal weights, and analyzed heat transfer performance using the field synergy principle [21]. It was demonstrated that the optimized structures with more branched channels could reduce temperature due to the augmentation of heat transfer area. However, this simultaneously weakened the synergy between the velocity and temperature fields, i.e., increased the synergy angle, since the flow branches lengthened flow path and expanded flow deterioration areas. Lin et al. [22] employed a pseudo-3D two-layer TO model [23] to account for the temperature distribution of the solid substrate, and optimized the average temperature and pumping power, leading to a reduction in temperature variance of up to 155.8% and an 89% improvement in the performance evaluation criterion (PEC) compared to the rectangular channel liquid cooling plate. Since most existing TO studies have been limited to laminar flow regime, Wang et al. [24, 25] conducted numerical and experimental investigations of turbulent TO methods and performance of the optimized liquid cooling plates. Compared to the conventional rectangular channel liquid cooling plate, the average temperature of the turbulent TO results was reduced by 6.8 K [24], and the PEC was improved by 8.2% compared to the laminar TO liquid cooling plate [25],.

The density-based TO method treats the design domain as a porous medium and porosity as the design variable [26], obtaining the designed liquid cooling plate structure by optimizing the porosity distribution. The foundational heat transfer principle for liquid cooling plate TO is the Newton's cooling law, i.e., convective heat transfer [27, 28], through which the thermal objective function should directly target the total convective heat transfer depicted as $h^*A\Delta T$, requiring both the heat transfer area and heat transfer coefficient to be explicitly formulated and updated iteratively. However, since the density-based TO method struggles to clearly capture solid-liquid interfaces, nor can it reflect the influence of other complex topological features beyond porosity. Furthermore, structural topology of fins and flow channels of TO liquid cooling plate is highly complex, and they continuously evolve within each iteration, leading to iteratively varying velocity and temperature fields. Consequently, it is almost impossible to explicitly obtain and iteratively update the heat transfer coefficient of the optimizing liquid cooling plate using any empirical or semi-empirical formula during the TO process. Therefore, most of the previous TO studies simplified the generalized heat transfer coefficient as a constant, while simplifying the heat transfer area to be linearly proportional to the solid fraction, leading to a largely simplified thermal objective function in the most common form of $\phi_t(\gamma)=h^*(1-\gamma)(T_r-T)$ [20, 29, 30], which hardly reflect the mechanism of convective heat transfer.

Recently, Han et al. [31, 32] proposed a fractal geometry topology optimization (FGTO) method that incorporates fractal geometry theory into the density-based TO framework, which explicitly describes the evolving scaling of heat transfer area with respect to design variables during the TO process through an additional design freedom of fractal dimension. It was demonstrated that this method can more effectively reflect the characteristics of convective heat transfer in the objective function, thereby enabling more complex structural topologies, improved heat transfer areas, and superior performance in the optimized liquid cooling plates under both laminar and turbulent flow conditions, compared to the conventional TO methods. However, this method still treats the heat transfer coefficient as a constant and considers only the effect of the heat transfer area. Therefore, it does not yet provide a comprehensive explicit depiction and direct optimization of convective heat transfer. Several approaches have been attempted to incorporate a varying heat transfer coefficient into density-based TO framework. Lad et al. [33] derived an empirical relationship between the heat transfer coefficient and magnitude of velocity under the conventional rectangular liquid cooling plate through 3D numerical simulations, and directly incorporated this relationship into the TO to achieve a pseudo-3D TO with higher fidelity. However, this empirical relationship derived from conventional structures cannot effectively reflect the flow and heat transfer characteristics of other complex structural topologies in the TO liquid cooling plates. Yang et al. [34] attempted to calculate the interlayer heat transfer coefficient between the solid substrate layer and the fluid layer along the thickness direction in a multi-layer TO model by assuming that the ratio of heat transfer coefficient to the solid-phase thermal resistance is a constant. Such an approach does not account for the heat transfer coefficient on the lateral interfaces between fluid and solid fins, which in fact dominates the convective heat transfer in the optimized liquid cooling plates.

The field synergy theory [21, 35] is widely applied in performance evaluation of heat exchangers and provides a qualitative guidance for optimization [36, 37]. The synergy angle can be used to evaluate the heat transfer performance, and the synergy number enables to intrinsically estimate the heat transfer coefficient based on the velocity and temperature fields, unlike empirical equations that rely on specific channel and fin geometries. Therefore, the field synergy theory is inherently compatible with the liquid cooling plate TO, and some studies have incorporated it to evaluate and analysis the heat transfer performance of the TO results [20]. However, since the field synergy theory intrinsically describes the heat transfer of the fluid, it is incapable of directly incorporating the influence of heat transfer area and solid distribution, leading to the failure of directly applied to solid-fluid conjugated optimization problem. Therefore, most current applications of the field synergy theory only serve primarily as a powerful principle for performance evaluation rather than a direct tool for design or optimization of the heat exchanger, etc. Marti et al. [38] attempted to perform TO singly targeting the synergy angle, but the optimized structures were demonstrated to be very simple and even some impractical results emerged in the optimization results. Consequently, simply targeting the synergy angle or synergy number as the objective function fails to achieve effective convective heat transfer optimization. Since the FGTO method enables explicit depiction of lateral heat transfer area, while the field synergy theory can reflect the effect of varying velocity and temperature fields on the heat transfer coefficient, it is very natural to integrate those two parameters together to achieve a comprehensive explicit description and direct optimization of convective heat transfer.

In this study, a convective heat transfer topology optimization (CTO) method developed from the FGTO method is proposed, where the heat transfer coefficient is described by the field synergy theory, and the heat transfer area is explicitly described by the fractal geometry theory to achieve a direct optimization of convective heat transfer through the thermal objective function. The methodology of the CTO framework is introduced in **section 2.1-2.3**, with the application of field synergy theory discussed in **section 2.1**, the explicit description of heat transfer area by the FGTO method provided in **section 2.2**, and the basic control equations for laminar and turbulent TO frameworks provided in **section 2.3**. The CTO is conducted in both the laminar and turbulent flow regimes (abbreviated as Laminar-CTO and Turbulent-CTO), and the optimization results are compared with the corresponding FGTO results without incorporation of field synergy theory (abbreviated as Laminar-FGTO and Turbulent-FGTO), with both the laminar and turbulent optimization results discussed in **section 3.1**. The 3D numerical calculations are conducted for performance evaluation, with performance under laminar flow conditions discussed in **section 3.2**, while that under turbulent flow conditions discussed in **section 3.3**. The mechanisms and differences of the effects of incorporating the field synergy theory into Laminar-FGTO and Turbulent-FGTO are further investigated, which is also discussed in **section 3.3**. Furthermore, a prototype of the optimized liquid cooling plate is fabricated and the manufacturability and feasibility of the proposed CTO method is discussed in **section 3.4**. It is demonstrated that the CTO enhances the synergy of optimization across different regions and enables more proactive circumvention of local optimal solutions, generating more directional structural topology, compared to the FGTO. Consequently, it reduces the stagnation zones with degraded flow and heat transfer performances, improves the flow direction in branched channels, leading to an improved heat transfer coefficient and thermal-hydraulic performances. The CTO method provides a paradigm that directly applies field synergy principle to liquid cooling plate design through integrating it with the FGTO framework according to the fundamental mechanisms of convective heat transfer, which not only refines the liquid cooling plate TO framework but can also be extended to the optimization of other heat transfer enhancement devices.

## 2. CTO Model

### 2.1 Field synergy theory

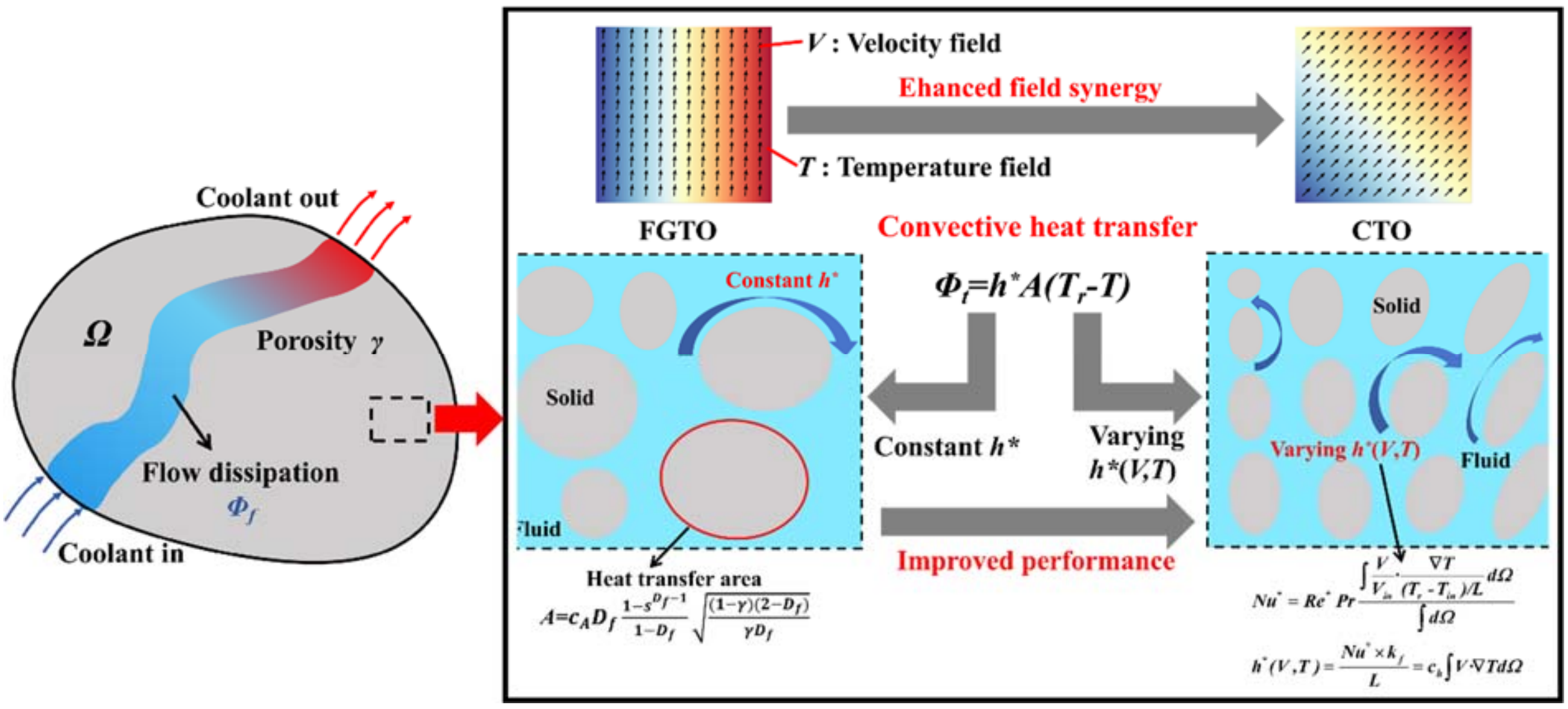

**Fig. 1.** Schematic of the CTO method, which incorporates the fractal geometry theory and field synergy theory into density-based TO to explicitly describe and optimize convective heat transfer through the objective function. The heat transfer area is depicted by the fractal geometry theory, while the heat transfer coefficient is described by the field synergy theory to explicitly reflect the effect of varying velocity and temperature fields on convective heat transfer

**Fig. 1** illustrates the schematic of the difference between the FGTO and CTO methods. Both the FGTO and CTO target the pumping power and convective heat transfer across the lateral heat transfer area explicitly depicted by the fractal geometry theory. The FGTO simplifies the heat transfer coefficient as a constant during TO process, since the continuously evolving structure, velocity and temperature fields make it difficult to explicitly describe the heat transfer coefficient. The CTO further improves the FGTO method by the field synergy theory, through which the iteratively evolving heat transfer coefficient can be explicitly depicted, as the field synergy theory enables the description of heat transfer coefficients directly based on the flow and temperature fields without relying on specific geometric structures. Consequently, the mechanism of convective heat transfer in liquid cooling plate can be conveyed more comprehensively, leading to an enhanced field synergy with superior performance achieved in the CTO results, compared with those by the FGTO. Technically, the CTO method proposed in this study represents a further development of the existing FGTO method, with incorporating both the field synergy and fractal geometry theories into TO framework, rather than the FGTO only adopted the fractal geometry theory.

The CTO and FGTO are conducted using the density-based TO method, which treats the design domain as porous media and porosity $\gamma$ as the design variable. Since the field synergy theory has also been widely adopted to describe the heat transfer in porous medium [39-41], it is fundamentally adaptable to be integrated with the density-based TO framework.

According to the field synergy theory, the average Nusselt number of the optimizing liquid cooling plate can be estimated based on the dimensionless velocity and temperature fields, which is formulated as [21, 40]:

$$Nu^{*}=Re^{*}Pr\frac{\int \frac{V}{V_{in}}\cdot\frac{\nabla T}{\frac{T_r-T_{in}}{L}}d\Omega}{\int d\Omega}=\frac{1}{A_p}\frac{\rho_f V_{in}L}{\mu}\times\frac{\mu C_{pf}}{k_f}\int\frac{V}{V_{in}}\cdot\frac{\nabla T}{\frac{T_r-T_{in}}{L}}d\Omega \tag{1}$$

where $A_p$ is the area of design domain, $L$ is the characteristic length for 2D TO calculation, which is adopted as the inlet width in this study, $V_{in}$ and $T_{in}$ are the inlet velocity and temperature, respectively, $\Omega$ is the design domain and $T_r$ is the reference temperature of the heat source. It should be noted that the characteristic lengths used in 2D TO and 3D numerical calculations are different, with the former being the inlet width $L$, while the latter being the inlet hydraulic diameter $D_h$ that incorporates the thickness of the liquid cooling plate. Consequently, the Reynolds number and Nusselt number used in 2D TO are denoted as $Re^{*}$ and $Nu^{*}$, respectively, to distinguish them from the $Re$ and $Nu$ used in 3D numerical calculations.

The average heat transfer coefficient is then obtained from the above $Nu^{*}$ as:

$$h^{*}(V,T)=\frac{Nu^{*}\times k_f}{L}=c_h\int V\cdot\nabla T d\Omega \tag{2}$$

where $c_h=\frac{\rho_f C_{pf} L}{(T_r-T_{in})A_p}$ is a constant independent of the design and state variables, so it can be eliminated in the dimensionless objective function in Eq. (7). It is demonstrated that the expression of the heat transfer coefficient in Eq. (2) is based on the physical essence of convective heat transfer and can be considered independent of any prior knowledge concerning the geometric structure. Therefore, such an approach is inherently compatible with TO framework, in which the fin and channel structures continuously evolve with iteration steps. It should be noted that Eqs. (1) and (2) are directly obtained from the previous studies, while the theoretical contribution of this study lies in proposing the concept of convective heat transfer topology optimization (CTO) by integrating field synergy and fractal geometry theories to describe the physical mechanism in liquid cooling plate. Furthermore, the engineering application scope of field synergy theory is expanded by integrating with the fractal geometry theory, so that it can serve not only as a criterion for performance evaluation but can also be directly applied for TO design.

The heat transfer coefficient is expressed in terms of an integral related to the flow and temperature fields by the field synergy theory, which implies there should be two distinct approaches to incorporating it into the TO framework: one involves a spatially varying local heat transfer coefficient, where the integration range is the smallest computational unit, i.e., the mesh grid; the other involves using a uniform average heat transfer coefficient over the design domain, with the integration range corresponding to the entire optimizing liquid cooling plate. Although the former approach undoubtedly aligns better with physical reality, the average heat transfer coefficient, rather than the local heat transfer coefficient, is adopted for the CTO, as formulated in Eqs. (1) and (2), which aims to provide a well-balance between the computational stability and physical essence.

On the one hand, a spatially varying heat transfer coefficient in the objective function could cause high nonlinear coupling, which would lead to prohibitively high computational costs and computational instability in the sensitivity analysis. On the other hand, since both the flow and temperature fields are spatially continuous, even a slight structural change in any region driven by the objective function would inevitably affect the field synergy in other regions over the design domain. Therefore, the entire design domain should be directly optimized as a unified entity in the thermal objective, rather than attempting to indirectly improve overall performance by targeting local heat transfer, which might lead to various design cells ineffectively competing each other and easily getting trapped in local optimal solutions or even failing to converge. Consequently, the average heat transfer coefficient rather than spatially varying local heat transfer coefficient is adopted in the CTO, to ensure that the updates of design variables throughout the design domain contribute to the overall heat transfer performance of the optimized liquid cooling plate, rather than focusing solely on the optimization of a few isolated areas.

Furthermore, most TO studies adhere to a workflow of low-fidelity optimization followed by high-fidelity validation, to mitigate the issue of convergence difficulties, computational instability and performance degradation of optimization occurring under high-fidelity conditions [26, 42, 43]. Within such a TO framework, a comprehensive and physically consistent description of the interactions and evolving scaling relationships among various physical mechanisms in convective heat transfer is even more critical for obtaining high-performance results, compared to pursuit absolute quantitative agreement with high-fidelity results for single physical quantity. Despite using an average heat transfer coefficient rather than a locally varying heat transfer coefficient in Eq. (2), this approach still ensures a physically consistent scaling relationship between the heat transfer coefficient and the velocity and temperature fields. Therefore, this simplification not only demonstrates numerical feasibility but also possesses physical plausibility within the TO framework.

## 2.2 Objective function

The topological features beyond the porosity are unified by the fractal dimension $D_f$ , which is defined by the fractal geometry theory and formulated as [44]:

$$D_f=2+\frac{ln\,(1\text{-}\gamma)}{ln\,(s)} \tag{3}$$

where $s$ is an input parameter related to the fractal dimension in the CTO, which represents the ratio of the maximum pore size to the minimum one. According to [31, 32], $s$=8000 is recommended to achieve the optimal thermal performance and PEC while ensuring computational convergence and stability, under both laminar and turbulent flow conditions.

According to our previous study [31], the lateral heat transfer area of the porous structure per unit volume, i.e., the specific area, of the optimizing porous structures is explicitly depicted by the fractal geometry theory and can be formulated as:

$$A=c_A D_f\frac{1\text{-}s^{D_f\text{-}1}}{1\text{-}D_f}\sqrt{\frac{1\text{-}\gamma}{\frac{D_f}{2\text{-}D_f}\gamma}} \tag{4}$$

where $c_A$ is a dimensionless constant representing shape parameter, and it can be eliminated in the dimensionless objective function in Eq. (7) and thus will not affect the TO process.

The thermal objective function that characterizes the total convective heat transfer is formulated as:

$$\Psi_t=\int_\Omega \phi_t(\gamma)\,d\Omega=\int_\Omega h^*(V,T)A(T_r\text{-}T)\,d\Omega \tag{5}$$

where $\phi_t(\gamma)$ is the local thermal objective function, $\Omega$ is the design domain region, and $T_r$ is the reference temperature of the heat source. For the CTO method, the heat transfer coefficient $h^*(V,T)$ is estimated by the field synergy theory in Eq. (2). For the FGTO method, it is simplified as a constant so it can be eliminated in the dimensionless function in Eq. (7).

The total flow dissipation rate, i.e., the pumping power, is chosen as the hydraulic objective function to balance flow resistance of the optimized liquid cooling plates [45]:

$$\Psi_f=\int_\Omega \phi_f(\gamma)\,d\Omega=\int_\Omega (\mu\,\nabla V\cdot\nabla V+f_p\cdot V)d\Omega \tag{6}$$

where $\phi_f(\gamma)$ is the local flow dissipation and $f_p$ is the flow resistance defined in Eqs. (10) and (11). The thermal and hydraulic objective functions are normalized and weighted to obtain the total objective function as follows:

$$\Psi=w_t\Psi^*_t-w_f\Psi^*_f \tag{7}$$

where $w_t$ and $w_f$ are the weights of the dimensionless thermal objective $\Psi^*_t=\frac{\Psi_t}{\Psi_{t0}}$ and dimensionless hydraulic objective $\Psi^*_f=\frac{\Psi_f}{\Psi_{f0}}$, respectively, with $w_t+w_f=1$. $\Psi_{t0}$ and $\Psi_{f0}$ are the initial values of thermal and hydraulic objectives, respectively. The subscripts t and f in objective function represent thermal and hydraulic objectives, respectively.

Although the initial conditions of design variable might affect geometric details of the optimized structures, the effectiveness of the CTO method and the underlying mechanisms all revolve around the fundamental physical mechanism of convective heat transfer, which is universally suitable without relying on specific initial structures. Consequently, the initial condition of the design variable field is set as a uniform distribution with $\gamma$=0.5 in this study, which is normally adopted in other TO studies [12, 18].

## 2.3 Laminar and turbulent TO models

The CTO and FGTO are conducted and compared under both the laminar and turbulent flow conditions. The fluid in the design domain is assumed to be incompressible flow and the continuity and momentum conservation equations are expressed as:

$$\nabla\cdot\left(\rho_f V\right)=0 \tag{8}$$

$$\nabla\cdot\left(\rho_f VV\right)=\nabla\cdot\left((\mu+\mu_T)(\nabla V+\nabla V^T)\right)-\nabla p-f_p \tag{9}$$

where $\mu_T=\rho C_\mu\frac{k^2}{\varepsilon}$ is the turbulent viscosity and it is zero in the laminar flow regime. The $f_p$ represents the flow resistance by porous media, for the Laminar-CTO and Laminar-FGTO optimized in laminar flow regime, it is depicted by the Darcy model formulated as [46]:

$$f_{pl}=\frac{\mu}{DaL^2}\frac{q_a(1-\gamma)}{q_a+\gamma}V \tag{10}$$

where $Da$ is the Darcy number, $q_a$ is the penalty factor and $L$ is the characteristic length. For the Turbulent-CTO and Turbulent-FGTO optimized in the turbulent flow regime, a Forchheimer term that represents the nonlinear behavior of flows at high $Re$ is added, and the flow resistance term can be formulated as [47]:

$$f_{pt}=\left(\frac{\mu}{Dal^2}V+\frac{C_d\ \rho_f}{l\sqrt{Da}}|V|V\right)\frac{q_a(1-\gamma)}{q_a+\gamma} \tag{11}$$

where $C_d$ is the drag coefficient.

The energy conservation equation is expressed as [29]:

$$\gamma\rho_f C_{pf}(V\cdot\nabla)T-\left[(1-\gamma)k_s+\gamma k_f\right]\nabla\cdot(\nabla T)=(1-\gamma)Q \tag{12}$$

where $Q$ is the volumetric heat source and the subscripts $s$ and $f$ represent solid and fluid, respectively.

Additional equations for turbulence variables are required for the Turbulent-CTO and Turbulent-FGTO, and the $k$-$\varepsilon$ turbulence model is adopted for the turbulent TO, with the $k$-equation and $\varepsilon$-equation formulated as:

$$\rho_f(V\cdot\nabla)k=\nabla\cdot\left[\left(\mu+\frac{\mu_T}{\sigma_k}\right)\nabla k\right]+\mu_T[\nabla V:(\nabla V+(\nabla V)^T)]-\rho_f\varepsilon+g_k \tag{13}$$

$$\rho_f(V\cdot\nabla)\varepsilon=\nabla\cdot\left[\left(\mu+\frac{\mu_T}{\sigma_\varepsilon}\right)\nabla\varepsilon\right]+C_{1\varepsilon}\frac{\varepsilon}{k}+\mu_T[\nabla V:(\nabla V+(\nabla V)^T)]-C_{2\varepsilon}\rho_f\frac{\varepsilon^2}{k}+g_\varepsilon \tag{14}$$

where the parameters in Eqs. (13) and (14) are $C_{1\varepsilon}$=1.44, $C_{2\varepsilon}$=1.92, $\sigma_k$=1, and $\sigma_\varepsilon$=1.3 [25], respectively. The terms $g_k$ and $g_\varepsilon$ represent the generation rate of the turbulent kinetic energy and dissipation rate induced by the porous media [48, 49], respectively, which can be formulated as [32]:

$$g_k=q_a\frac{1-\gamma}{q_a+\gamma}\times\frac{C_{turb}\rho_f}{l\sqrt{Da}}|V|k \tag{15}$$

$$g_\varepsilon=q_a\frac{1-\gamma}{q_a+\gamma}\times\frac{C_{turb}C_{2\varepsilon}\rho_f}{l\sqrt{Da}}|V|\varepsilon \tag{16}$$

where $C_{turb}$ is a constant that is recommended to be 0.28 [50].

The Helmholtz-type density filter is used to alleviate the mesh dependency problems [51]:

$$\gamma_f=r^2\nabla^2\gamma_f+\gamma \tag{17}$$

where $\gamma_f$ is the filtered design variable and $r$ is the filter radius, which is set as 2.5 × mesh size for laminar TO, and 1.5 × mesh size for turbulent TO. To minimize the intermediate region where the design variable have not fully converged to pure solid or liquid, the hyperbolic tangent projection is adopted as [52]:

$$\gamma_p=\frac{\tanh(\beta(\gamma_f-\gamma_\beta))+\tanh(\beta\gamma_\beta)}{\tanh(\beta(1-\gamma_\beta))+\tanh(\beta\gamma_\beta)} \tag{18}$$

where $\beta$, $\gamma_\beta$ and $\gamma_p$ are the projection slope, projection point and projected design variable, respectively. The parameters used in this study are listed in **Table 1**.

**Table 1** Values of the important parameters.

| | | | |
|---|---|---|---|
| $q_a$ | 0.01 | $C_{2\varepsilon}$ | 1.92 |
| $Da$ | $10^{-6}$ | $C_d$ | 1.0 |

| $\beta$ | 8 | $C_{turb}$ | 0.28 |
|---|---|---|---|
| $\gamma_\beta$ | 0.5 | $\sigma_k$ | 1.0 |
| $L$ | 0.01[m] | $\sigma_\varepsilon$ | 1.3 |
| $T_r$ | 370 [K] | $s$ | 8000 |
| $C_{1\varepsilon}$ | 1.44 | | |

## 2.4 Geometry and boundary conditions

The effectiveness of the CTO method is validated using the liquid cooling plate models shown in **Fig. 2**. **Fig. 2(a)** illustrates the application scenario of the optimized liquid cooling plate, which is integrated into the liquid cooling system for thermal management of electronic components such as CPUs and GPUs within server racks in DCs. The 2D liquid cooling plate model is adopted for topology optimization, while the performance of the optimized liquid cooling plate is evaluated using 3D numerical calculations. In practical applications, various coolants and solid materials can be selected for liquid cooling plates, e.g., water, ethylene glycol, or propylene glycol solutions as coolants, and copper or aluminum alloys as solid materials. Without limiting the generality of the CTO method, copper is adopted as the solid material and water as the coolant for the representatives in this study, with the thermo-physical properties listed in **Table 2**.

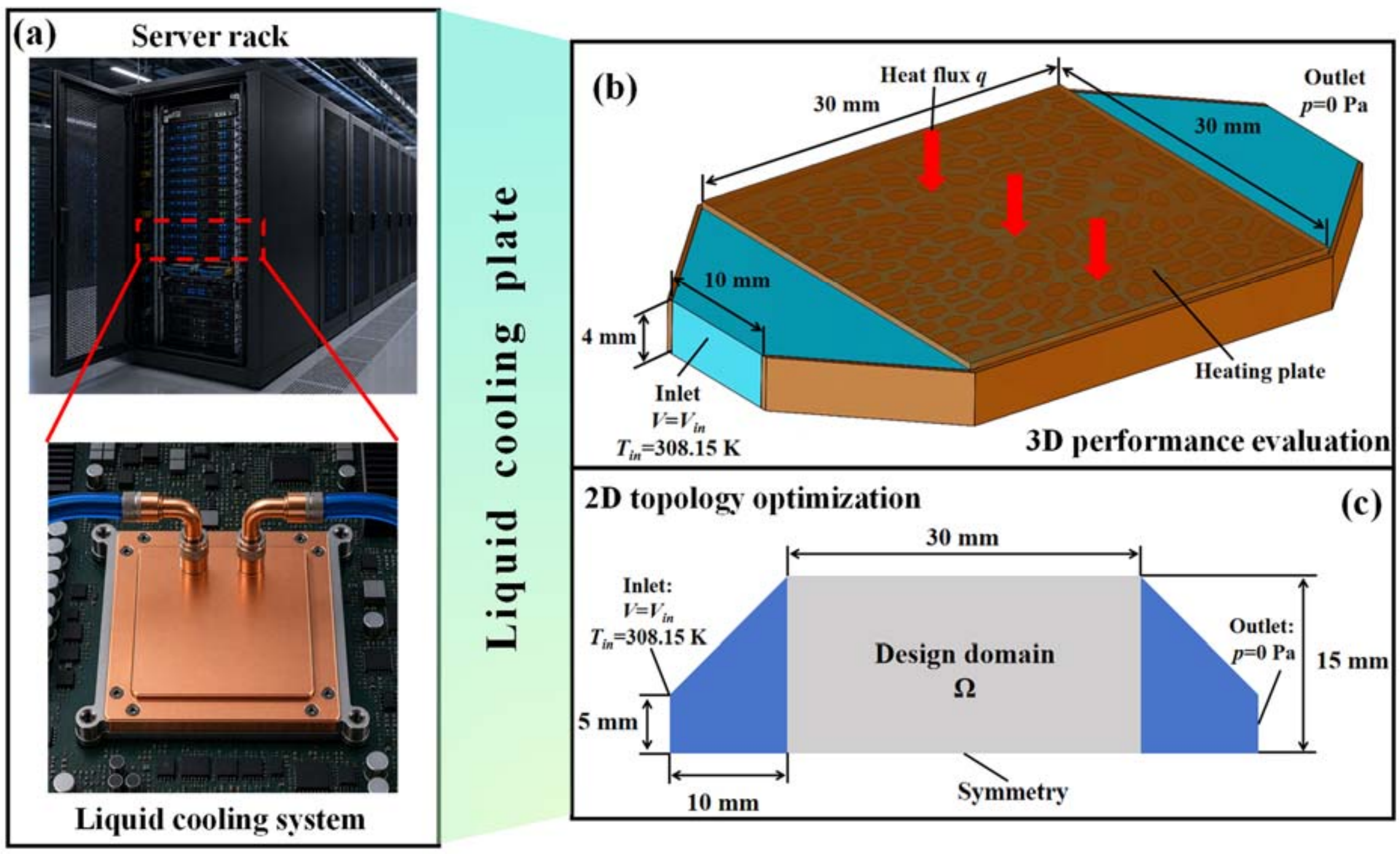

**Fig. 2.** Schematics of liquid cooling plate model. (a) schematic of the application scenario of liquid cooling

plate, (b) geometry structure and boundary conditions of the 3D numerical calculation model for performance evaluation, (c) schematic of the 2D topology optimization model

**Fig. 2(b)** shows the geometry and boundary conditions for the 3D liquid cooling plate model used in this study. The water flows into the liquid cooling plate with inlet velocity $V_{in}$, inlet temperature of 308.15 K, and $p$=0 Pa is set as the boundary condition of the outlet. In practical applications, the heat source distribution might differ depending on the device configuration. Considering that the heat source distribution would not significantly affect heat transfer performance under single-phase flow conditions, a uniform heat source is adopted to ensure the adaptability of this study to various heat source arrangement scenarios. As shown in **Fig. 2(b)**, a uniform heat flux $q$ is applied on the top surface of the heated plate, while all other walls are adiabatic. The 3D numerical calculations are conducted under both the laminar and turbulent flow conditions, as listed in **Table 3**. The $V_{in}$ ranges from 0.03-0.2 m/s with $q$=10 W/cm$^2$ under the laminar flow condition, and $V_{in}$ ranges from 0.7-3.5 m/s with $q$=100 W/cm$^2$ under the turbulent flow condition.

Due to the highly complex flow channel structure with significant variations in flow channel dimensions in the TO liquid cooling plates, the hydraulic diameter at the inlet $D_h$=2×4 mm×10 mm/(4 mm+10 mm)=5.7 mm is used as the characteristic length for estimating the Reynolds number in the 3D numerical calculations [9, 53]. Assuming that the coolant of water is an incompressible fluid with constant physical properties listed in **Table 2,** and using the inlet velocity $V_{in}$ as the characteristic velocity, the $Re=\rho_f V_{in} D_h/\mu$ of turbulent flow conditions is around 5,500–27,500, while the $Re$ of laminar flow conditions is around 240-1,570. It should be noted that the inlet velocity range is selected not only to ensure the flow remaining within the corresponding flow regimes, but also to prevent the heating surface temperature from becoming excessively high. For example, the average temperature of CPU/GPU cores is recommended to be controlled below 70 °C, otherwise the reliability would be compromised [13, 54]. Consequently, the lower bound of testing inlet velocity is set as 0.03 m/s for laminar flow condition, while as 0.7 m/s for turbulent flow condition to avoid an excessively low flow velocity and high temperature rise.

**Table 2** Thermo-physical properties of the solid and fluid.

| Material | $\rho$ [kg/m³] | $C_p$ [J/(kg·K)] | $k$ [W/(m·K)] | $\mu$ [Pa·s] |
|---|---|---|---|---|
| Water [55] | 993 | 4183 | 0.62 | 0.00072 |

| Copper | 8900 | 385 | 385 | - |
|---|---|---|---|---|

**Table 3** The operating conditions of the laminar TO and turbulent TO, and those of the 3D numerical calculations, including the laminar and turbulent flow conditions.

| Operating condition | 2D TO | | 3D numerical calculation | |
|---|---|---|---|---|
| | Laminar TO | Turbulent TO | Laminar flow | Turbulent flow |
| $q$ (W/cm$^2$) | 10 | 100 | 10 | 100 |
| $T_{in}$ (K) | 308.15 | 308.15 | 308.15 | 308.15 |
| $V_{in}$ (m/s) | 0.01 | 1.0 | 0.03-0.2 | 0.7-3.5 |

The CTO is performed using the 2D liquid cooling plate model, as shown in **Fig. 2(c)**, including the optimization in both the laminar (Laminar-CTO) and the turbulent (Turbulent-CTO) flow regimes. The Laminar-CTO and Turbulent-CTO results are compared with those by the Laminar-FGTO and Turbulent-FGTO methods, respectively, which are conducted under their corresponding flow conditions without incorporating field synergy theory, with all other parameters and operating conditions kept the same. The operating conditions used for the laminar TO and turbulent TO are listed in **Table 3**. The Laminar-CTO and Laminar-FGTO are conducted at $q$=10 W/cm$^2$ and $V_{in}$=0.01 m/s with a relatively low $Re$ to avoid the computational instability and poor convergence issues under high $Re$ conditions [26, 56]. The Turbulent-CTO and Turbulent-FGTO are conducted at $q$=100 W/cm$^2$ and $V_{in}$=1 m/s. Half of the liquid cooling plate with a symmetry boundary is employed to reduce computational costs, and all the other walls are adiabatic.

## 2.5 Data reduction

The Nusselt number $Nu$ is used to evaluate heat transfer performance of the optimized liquid cooling plates, which is formulated as:

$$Nu=\frac{h_{avg}D_h}{k_f} \tag{19}$$

where $h_{avg}=\frac{Q_{total}}{A_{eff}(T_{wall}-T_f)}$ is the average heat transfer coefficient [18], $Q_{total}$ the total heat applied to the top surface of the heated plate, $A_{eff}$ the effective heat transfer area including the lateral interface between fluid and solid fins, and the top interface between the fluid and top heating plate. The $T_f=\frac{T_{in}+T_{out}}{2}$ is the average temperature of fluid, $T_{wall}$ the average temperature of solid wall corresponding to the $A_{eff}$, respectively. The hydraulic diameter $D_h$ is estimated based on the dimensions at the inlet [9, 53], due to the significant variations in flow channel dimensions in TO liquid cooling plates.

The Nusselt number represented by the projected area (i.e., top heated surface area or the design domain area) [25, 57], is used to uniformly compare thermal performance of various liquid cooling plates with different heat transfer areas, which is formulated as:

$$Nu_p=\frac{Nu\times A_{eff}}{A_p} \tag{20}$$

The performance evaluation criterion (PEC) is adopted for evaluating overall thermal-hydraulic performance of the optimized liquid cooling plates, which is formulated as [58, 59]:

$$\mathrm{PEC}=\frac{Nu_p/Nu_{pr}}{(\Delta P/\Delta P_r)^{1/3}} \tag{21}$$

Where, subscript r refers to the reference benchmark of the pin fin liquid cooling plate. A higher PEC value indicates a better overall performance. and PEC>1 implies performance superior to that of the pin fin liquid cooling plate.

The synergy angle is used to evaluate synergy between flow and temperature fields of various liquid cooling plates. A smaller synergy angle indicates better synergy between the flow and temperature fields. The synergy angle is expressed as [35, 60]:

$$\theta=\frac{\int arccos\left(\frac{|V\cdot \nabla T|}{|V||\nabla T|}\right)d\Omega}{\int d\Omega} \tag{22}$$

**2.6 Numerical methods**

The CTO is solved in COMSOL Multiphysics (6.0). The design variables are firstly initialized with $\gamma$=0.5, and the state variables are calculated by the finite element method. Then the design variable is updated by the Global Convergence Method of Moving Asymptotes optimization solver (GCMMA) [61]. The optimization is terminated when the maximum residual of the design variable is less than $10^{-4}$. The 3D numerical calculations are further solved using the second-order upwind scheme for the discretization of momentum, energy, and turbulence variables equations, while a second-order discretization scheme is adopted for the pressure equation. The calculation is terminated when the residual of mass is less than $10^{-5}$, residuals of velocity and turbulence variables are less than $10^{-6}$, while the residual of energy is less than $10^{-8}$.

## 2.7 Mesh-independent analysis

The structured square mesh is adopted for the CTO calculation. Since the required grid number for the optimization increases with the structural complexity, the Laminar-CTO at $w_t$=0.7 with the most complex structural topology is used for the mesh-independent check representatively. The deviation in the objective function becomes insignificant when the grid number exceeds 45,247, as shown in **Table 4**. Therefore, considering both computational efficiency and accuracy, 45,247 mesh grids are used for the CTO calculations.

**Table 4** The results of the 2D mesh-independent check.

| Grid number | $\Psi$ | Deviation |
|---|---|---|
| 6614 | 0.4023 | 27.69% |
| 13,225 | 0.5137 | 8.47% |
| 24,173 | 0.5572 | 2.17% |
| 45,247 | 0.5693 | 0.37% |
| 78,511 | 0.5714 | - |

As for the 3D numerical calculations, the structured hexahedral mesh is adopted for the solid region and the unstructured tetrahedral mesh is adopted for the fluid region. The boundary mesh refinement is adopted with first layer thickness of 0.01 mm to improve the calculation accuracy near the solid-liquid interface. The mesh-independent check is conducted using the Laminar-CTO liquid cooling plate at $w_t$=0.7 with the most complex structural topologies under the turbulent flow condition of $V_{in}$=3.5 m/s, $T_{in}$=308.15 K and $q$=100 W/cm$^2$. As shown in **Table S1**, the deviations become unremarkable when the mesh grid number is 4.72 million, with deviations of $T_m$ and $\Delta P$ reduced to 0.72% and 0.05%, respectively. Consequently, a grid number of 4.72 million is ultimately adopted for the 3D numerical calculations, with corresponding mesh grid structure shown in **Fig. S1**.

## 2.8 Verification of the 3D numerical calculation

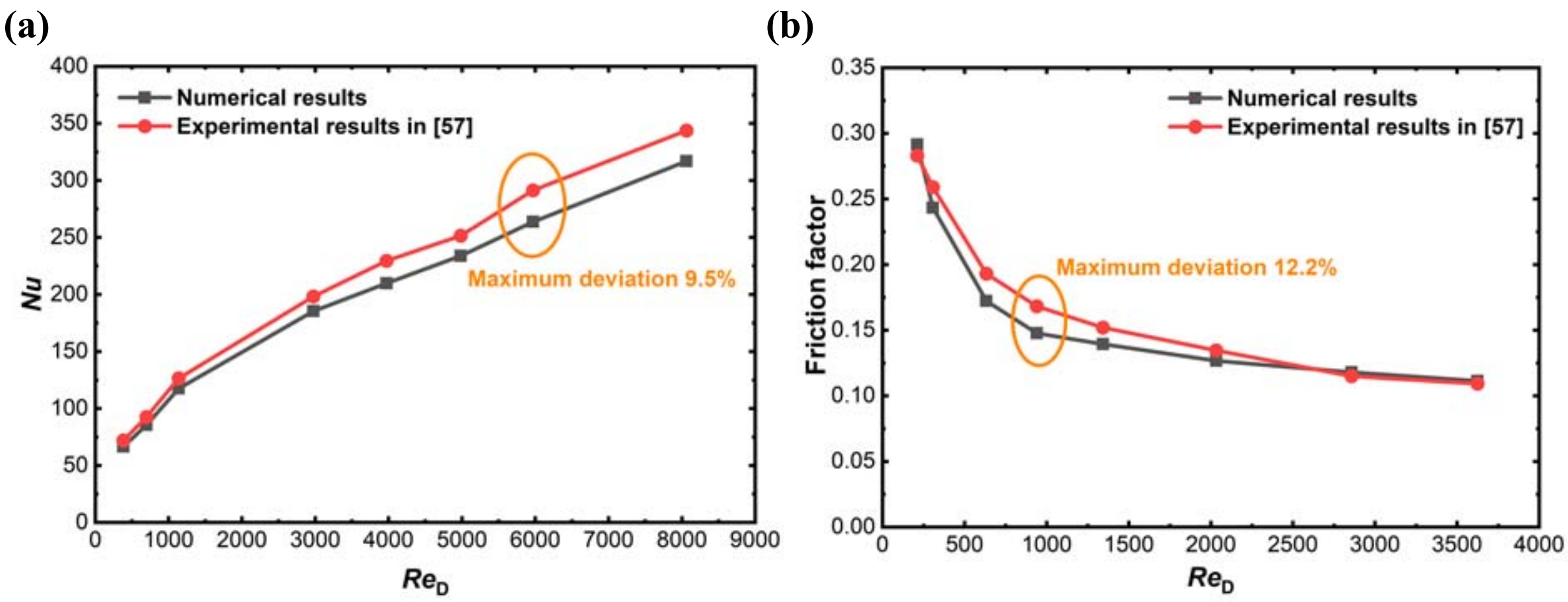

**Fig. 3.** Comparison of the numerical results and experimental results in [57]. (a) variation of the Nusselt number with the Reynolds number, (b) variation of the friction factor with the Reynolds number

The 3D numerical calculation model is verified by the experimental results in [57]. The verification is operated under a pin fin liquid cooling plate, with fin diameter around 3.67 mm, transverse pinch of 5.0 mm, longitudinal pinch of 4.33 mm and height of 4.0 mm, whose detailed geometric structure and boundary conditions are the same as those in [57]. The original experiment was conducted using water as the coolant with inlet temperature around 27 °C. Approximately uniform heat fluxes ranging from 1.6-24.4 W/cm$^2$ that vary accordingly with the testing Reynolds number are applied to the base substrate with an area of 35.7 cm$^2$. **Fig. 3** shows the comparison of the numerical results and experimental results, with the Reynolds number $Re_D$ based on the diameter of pin fin ranging from 210 to 8,060. It can be observed that the maximum deviation between the experimental and numerical results for the *Nu* and friction factor are 9.5% and 12.2%, respectively. Therefore, the 3D numerical calculation can be considered reliable.

## 3. Results and discussion

### 3.1 Effects of field synergy on the optimized liquid cooling plates

**Fig. 4(a)** and **Fig. 4(b)** show the structural topologies of the Laminar-CTO and Laminar-FGTO liquid cooling plates at various $w_t$, respectively. It is apparent that the incorporation of field synergy demonstrates a significant impact on the optimization results, with the solid fins in the Laminar-CTO results exhibiting greater selective orientation and a more hierarchically structured overall arrangement, compared to the Laminar-FGTO results. In the Laminar-CTO results, the solid near the inlet extends in the direction of the incoming flow, as shown by the red square in **Fig. 4(a)**, and is interspersed with some more intricate fins to achieve a more rational flow distribution while reducing local pressure drop. Since the fluid is heated along the main flow direction, this oriented solid structure facilitates better synergy between the temperature gradient and local velocity direction. In contrast, in the Laminar-FGTO results, the solid near the inlet is generated along the design domain boundary that is perpendicular to the flow direction, as shown by the red square in **Fig. 4(b)**, dispersing the incoming flow into various branches through collision in a coarser manner, which makes it more difficult to achieve a reasonable flow distribution while increasing local pressure drop.

As shown by the red circle in **Fig. 4(b)**, the branched channels in the downstream region of the Laminar-FGTO liquid cooling plates extend nearly perpendicular to the main flow or even slightly countercurrent, which is detrimental to efficient heat removal. In contrast, as shown by the red circle in **Fig. 4(a)**, the branched channels in the Laminar-CTO results exhibit more directional structural topologies which extend in a direction aligning to the main flow, leading to a better synergy and more efficient heat dissipation. Furthermore, in the regions on either side of the liquid cooling plate where flow and heat transfer are compromised, the solid in the Laminar-CTO results is more fragmented than those in the Laminar-FGTO results, which enhances local flow perturbations and disrupts the continuous boundary layer to improve heat transfer performance.

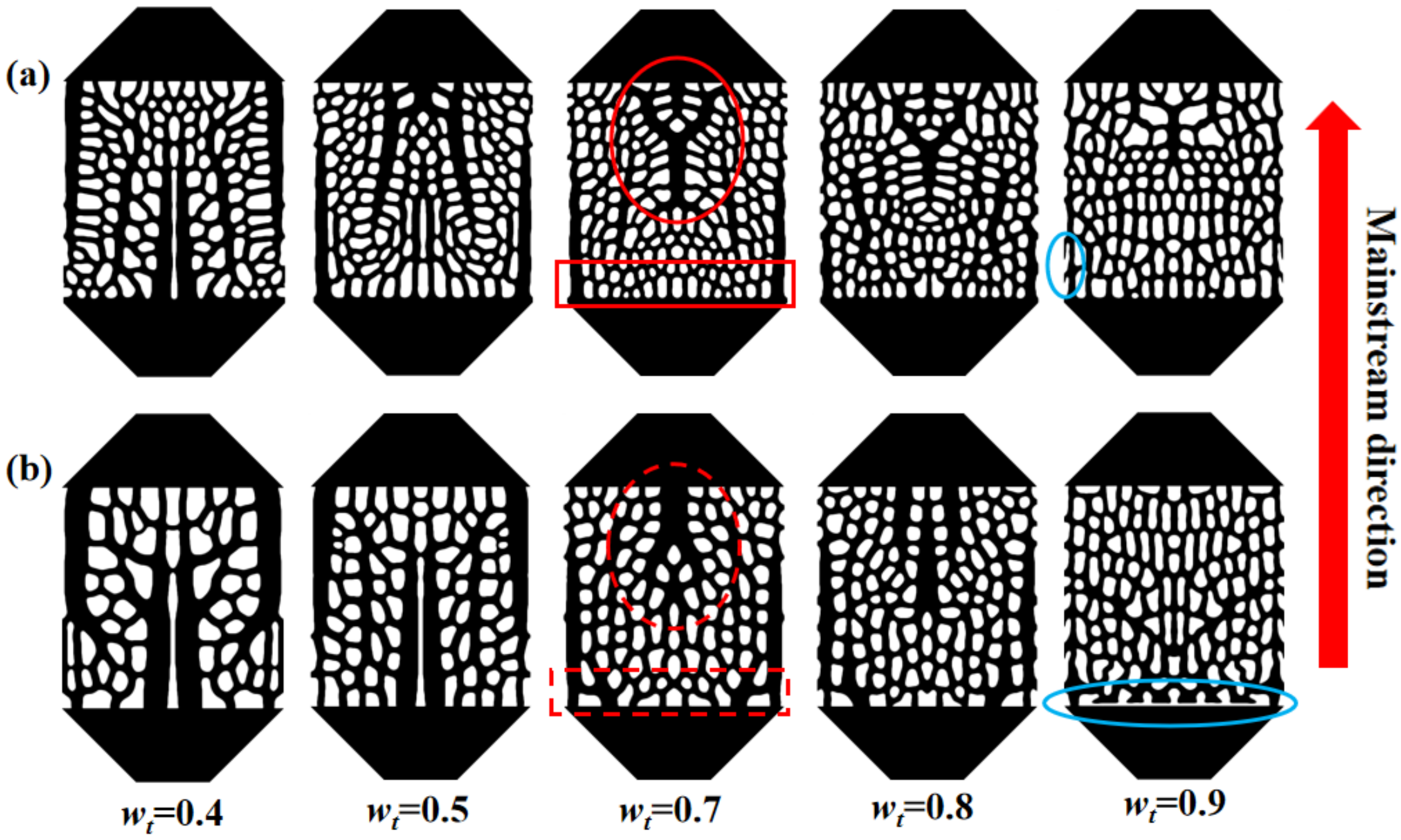

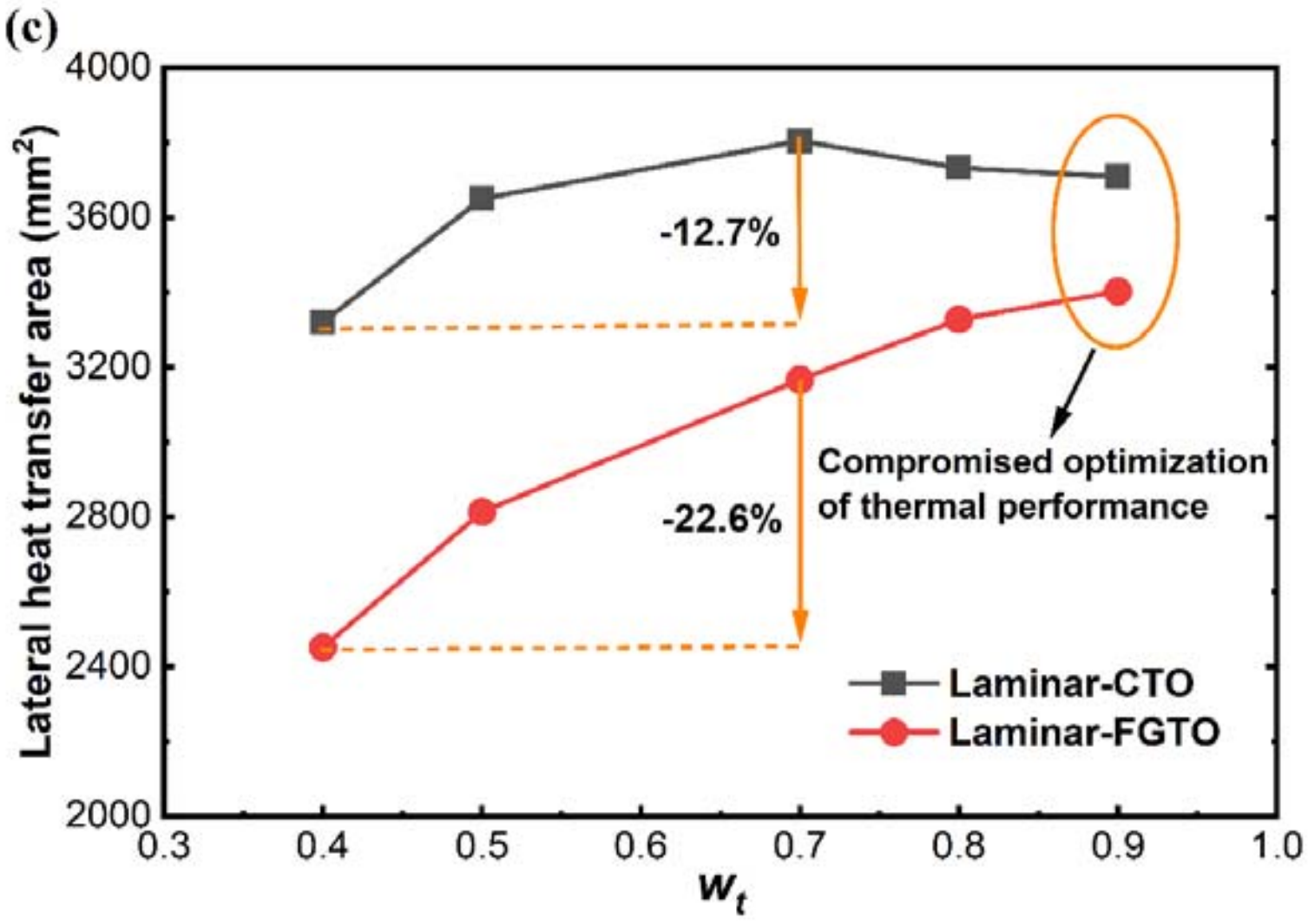

**Fig. 4.** The structural comparison of the optimized liquid cooling plates by the Laminar-CTO and Laminar-FGTO methods at various $w_t$. (a) the results by the Laminar-CTO method with the incorporation of field synergy theory, (b) the results by the Laminar-FGTO method without the incorporation of field synergy theory, (c) variation of lateral heat transfer area of various liquid cooling plates (thickness of 4 mm) with thermal weight $w_t$

According to the previous studies [31, 62], for multi-objective TO, excessively increasing or decreasing $w_t$ are not conducive to achieving a rational and high-performance structural topology, since it places excessive emphasis on one of the thermal-hydraulic objectives while neglecting the counterpart, leading to optimization more easily getting trapped in local optima. Although the CTO can more effectively balance the trade-off between thermal-hydraulic objectives due to the explicit consideration of the flow field in its thermal objective through field synergy theory, it can only mitigate the above issues rather than completely avoid them.

**Fig. 4(c)** shows the lateral heat transfer areas of the Laminar-CTO and Laminar-FGTO liquid cooling plates at various $w_t$. It is demonstrated that when the $w_t$ is decreased to a relatively small value of 0.4, the lateral heat transfer area for the Laminar-CTO and Laminar-FGTO results are reduced by 12.7% and 22.6%, respectively, compared to those at $w_t = 0.7$, indicating significantly compromised heat dissipation capabilities. As the $w_t$ is increased to an extremely large value of 0.9, some impractical structures like the flow dead zones can be observed in the Laminar-CTO results, as shown by the blue circles in **Fig. 4(a)**, and even blockage emerges in the Laminar-FGTO results, as shown by the blue circles in **Fig. 4(b)**. Furthermore, the improvement in heat transfer area gradually approaches saturation as the $w_t$ increases, leading to a compromised optimization effect on thermal performance at extremely large $w_t$. The lateral heat transfer area of the Laminar-FGTO results is augmented by 12.6% as the $w_t$ is increased from 0.5 to 0.7, while only by 7.4% as the $w_t$ is further increased from 0.7 to 0.9. Since the CTO should balance heat transfer area and field synergy, rather than simply optimizing heat transfer area as the FGTO, the lateral heat transfer area of the Laminar-CTO results even slightly decreases by 2.5% as the $w_t$ is increased from 0.7 to 0.9. Consequently, both the Laminar-CTO and Laminar-FGTO should avoid excessively increasing or decreasing $w_t$, and should appropriately adjust $w_t$ within the range around 0.5–0.8 to manipulate thermal-hydraulic performance.

According to the performance evaluation results in section 3.2, within the recommended $w_t$ range of 0.5–0.8, thermal performance can be improved by increasing $w_t$ at the cost of hydraulic performance, while the overall thermal-hydraulic performance indicated by the PEC exhibits minimal variation. For the Laminar-CTO results at $V_{in}$=0.1 m/s, $q$=10 W/cm$^2$ and $T_{in}$=308.15 K, the $Nu$ is improved by 21.3% and pressure drop is increased by 32.8% as the $w_t$ increases from 0.5 to 0.8, while the PEC is slightly improved by 7.8%, as shown in **Fig. 9**. For practical engineering applications, if there are no specific requirements regarding either thermal or hydraulic performance, the $w_t$ can be directly selected as approximately 0.8 to achieve an optimal PEC, or selected as around 0.7 to achieve the highest structural complexity with largest heat transfer area.

The significant difference in the structural topologies of the Laminar-CTO and Laminar-FGTO results stems from their fundamentally distinct optimization mechanisms driven by the incorporation of the field synergy theory. **Fig. 5(a)** and **Fig. 5(b)** show the iteration process of the Laminar-CTO and Laminar-FGTO at $w_t$=0.8, respectively, where the optimal thermal performance and PEC are achieved for both the Laminar-CTO and Laminar-FGTO. **Fig. 6(a)** and **Fig. 6(b)** show the evolution of flow and temperature fields of the Laminar-CTO and Laminar-FGTO liquid cooling plates, respectively. Additionally, the velocity and temperature gradient vectors are also provided in **Fig. 6**, to clearly demonstrate the synergy between velocity and temperature fields. **Fig. 7(a)** shows the variation of average design variable (liquid fraction) with iteration in the upstream one-third region in the optimizing liquid cooling plate. During the initial stage within 10 iteration steps, overall liquid fraction near the inlet rises rapidly with the formation of manifolds in both the Laminar-CTO and Laminar-FGTO results to reduce resistance to facilitate flow distribution, as shown in **Fig. 7(a)**. However, after this short initial stage, the Laminar-CTO and Laminar-FGTO exhibit significant differences in both the flow channel and solid fin evolutions. It can be observed from **Fig. 6(b)** that the distribution pattern of main flow channels with high local velocity in the Laminar-FGTO results are rapidly established at around 30 steps and exhibits minimal variation in the subsequent iterations. In contrast, the main flow channels in the Laminar-CTO results exhibit a progressively developing pattern, gradually extending from upstream to downstream, as shown in **Fig. 6(a)** and the red circles in **Fig. 5(a)**. In the Laminar-FGTO results, the isotherms were rapidly distorted across the entire design domain, as shown in **Fig. 6(b)**, whereas variation in the temperature distribution pattern extends progressively from upstream to downstream in the Laminar-CTO results, as shown in **Fig. 6(a)**.

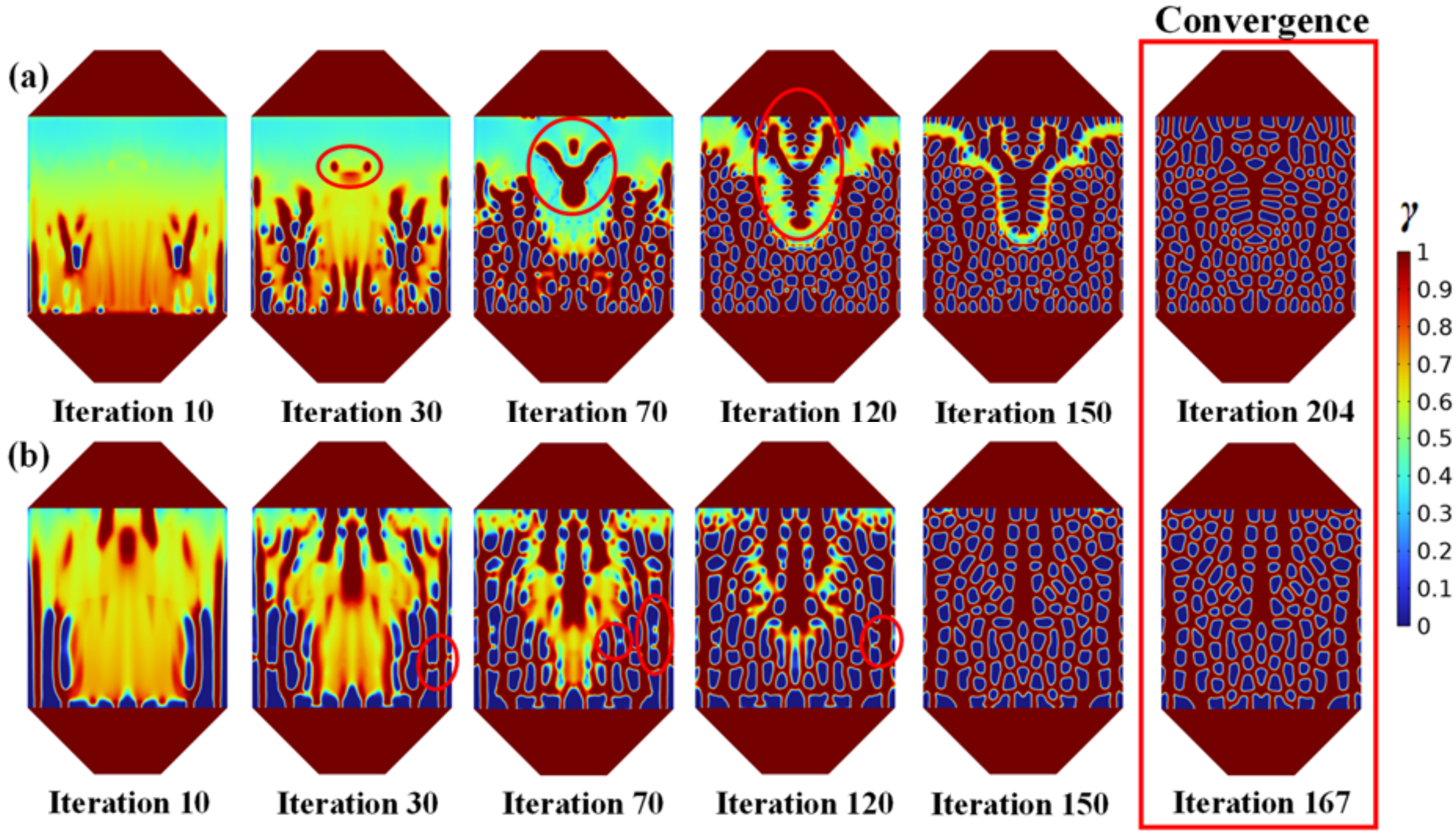

**Fig. 5.** Variation of the design variable fields of the Laminar-CTO and Laminar-FGTO results at $w_t$=0.8 with iteration steps. (a) the Laminar-CTO results, (b) the Laminar-FGTO results

The structural evolution extends gradually from the upstream to the downstream region, which can be attributed to two reasons: On the one hand, flow distribution, which affects the overall thermal-hydraulic performance of the entire liquid cooling plate, depends on the manifold structures in the upstream regions, therefore, the optimization naturally prioritizes the upstream region. On the other hand, the heat transfer coefficient is inherently higher upstream due to the inlet effect, which is demonstrated by the denser distribution of isotherms in the upstream region, as shown in **Fig. 6**, and further confirmed by the subsequent 3D numerical simulation results shown in **Fig. 12(b)**. Therefore, structural change upstream exhibits a greater impact on the optimization of heat transfer coefficient compared to that in the downstream region. Consequently, the Laminar-CTO, which explicitly includes the heat transfer coefficient in the objective function, tends to prioritize optimization of the upstream region first, before extending to the downstream region where the heat transfer coefficient varies less.

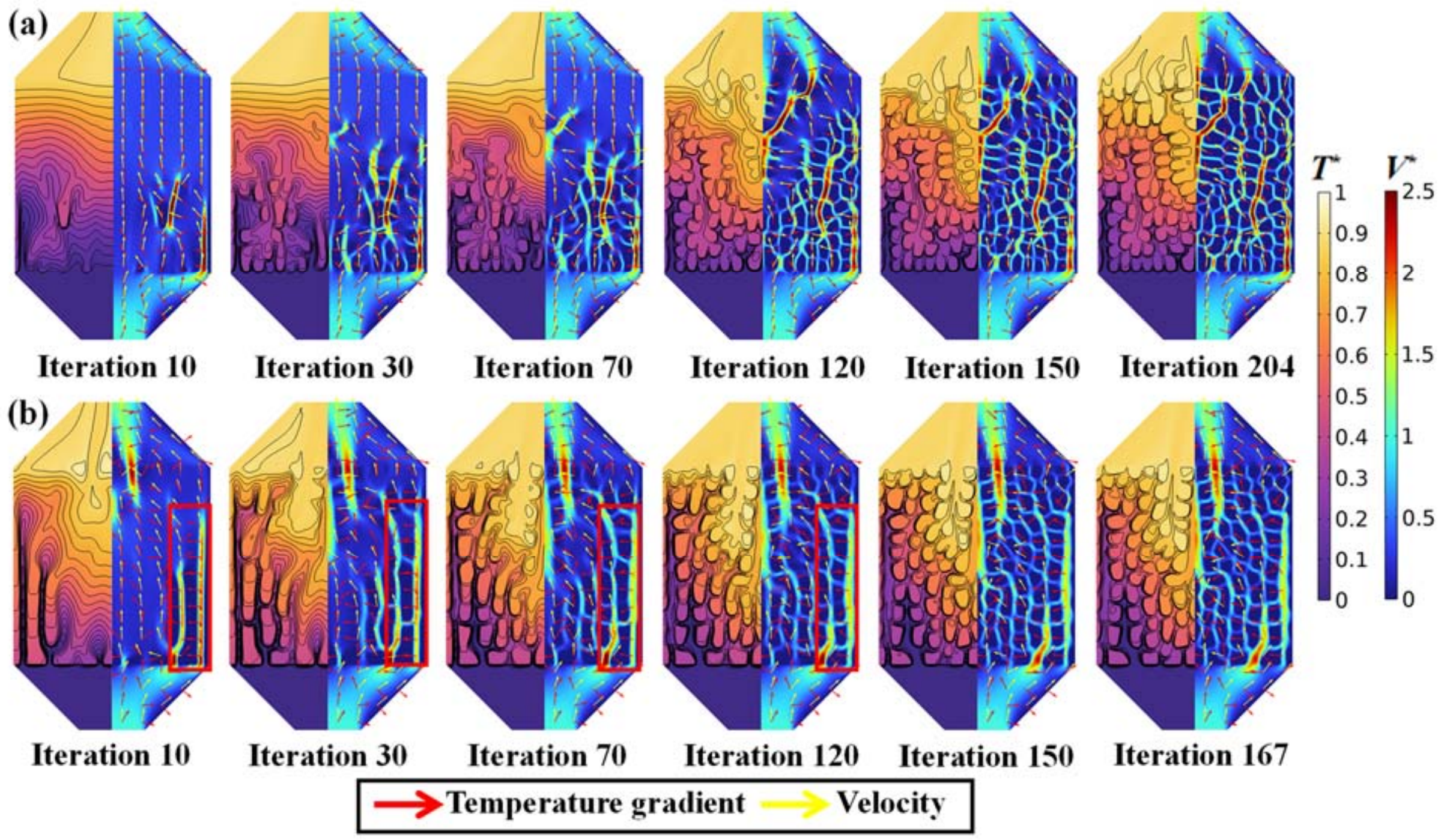

**Fig. 6.** Evolution of the dimensionless temperature $T^*=(T-T_{in})/(T_r-T_{in})$ and dimensionless velocity $V^*=V/V_{in}$ fields, and the velocity and temperature gradient vectors of the Laminar-CTO and Laminar-FGTO results at $w_t$=0.8 with iteration steps. (a) the Laminar-CTO results, (b) the Laminar-FGTO results

Different evolution patterns in flow and temperature fields are accompanied by distinct formation mechanisms of structural topologies: The Laminar-FGTO first builds large solid blocks during the early stage of the iteration, followed by gradual fragmentation as the iteration proceeds, as shown circled in **Fig. 5(b)**, whereas the Laminar-CTO can achieve fragmented solid fins directly. After the initial steps, in the upstream region, as shown in **Fig. 7(a)**, the liquid fraction of the Laminar-FGTO first drops by 20.1%, reaches a minimum after approximately 30 iterations, and then gradually rises by about 10.9% until convergence is achieved. This corresponds to the process that, large solid blocks first form rapidly, causing the liquid fraction to decrease, and then these blocks are iteratively broken up from which branched channels gradually emerge, as circled in **Fig. 5(b)**, leading to the increase in liquid fraction. In contrast, the liquid fraction in the Laminar-CTO continues to decrease by approximately 15.9% throughout the main iteration process without a significant inflection point, corresponding to the fragmented solid fins can be directly generated by Laminar-CTO, as shown in **Fig. 5(a)**. Since the fragmented structural topology is more conducive to disrupting the continuous thermal boundary layer and augmenting heat transfer area, while large solid block as local optima is detrimental to heat dissipation, it is demonstrated that the Laminar-FGTO first getting trapped in local optimal and then gradually escaping, while the Laminar-CTO can more effectively avoid the local optimal solutions.

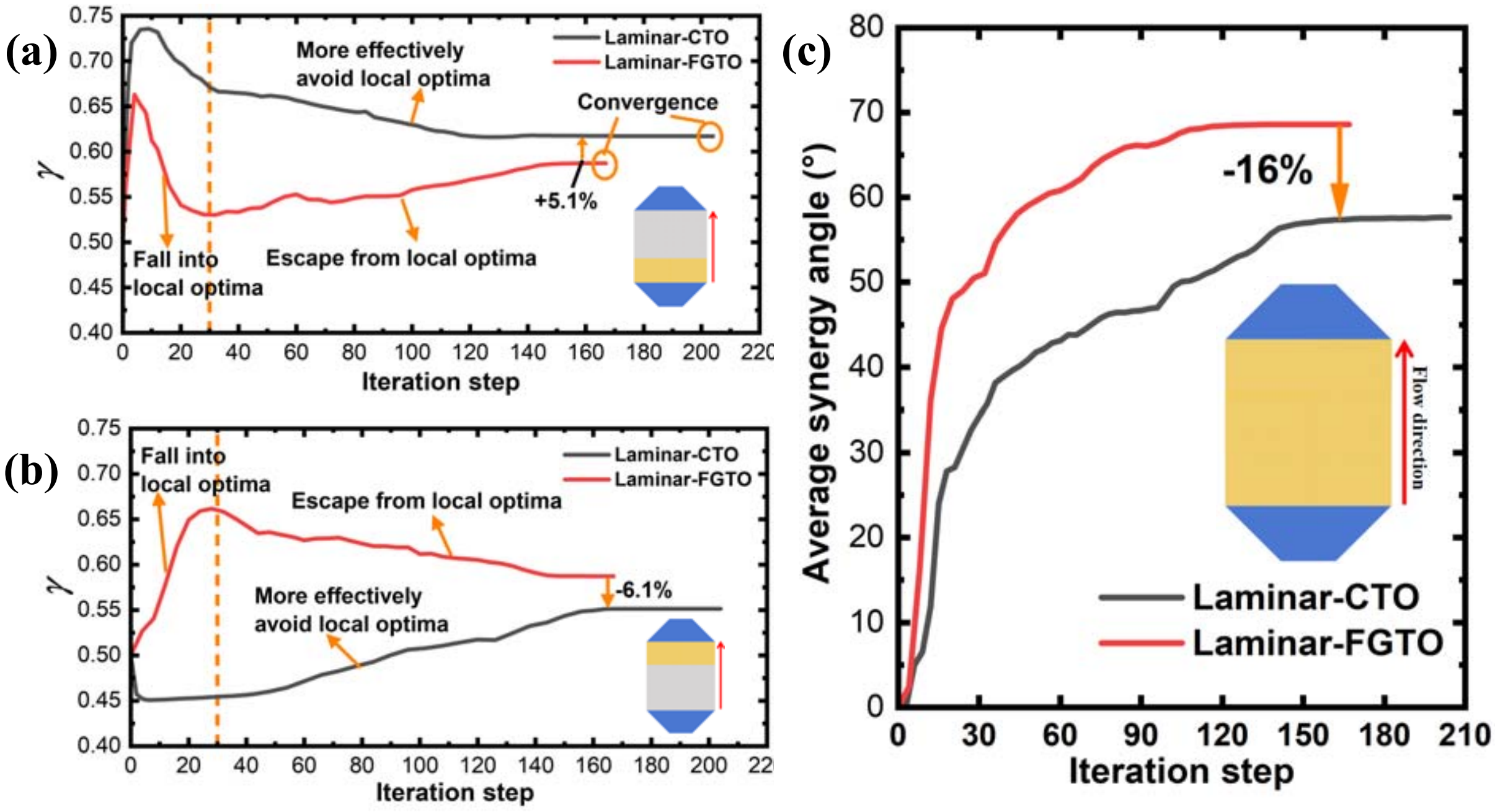

**Fig. 7.** Variations of the average design variable and synergy angle of the Laminar-CTO and Laminar-FGTO results at $w_t$=0.8 with iteration steps. (a) the average design variable of the upstream one-third region in the optimizing liquid cooling plate, (b) the average design variable of the downstream one-third region in the optimizing liquid cooling plate, (c) average synergy angle of the entire optimizing liquid cooling plate

**Fig. 7(b)** shows the variation of average design variable (liquid fraction) with iterations in the downstream one-third region in the optimizing liquid cooling plate. It can be observed that the liquid fraction in the downstream regions exhibit opposite variation trends to those in upstream region shown in **Fig. 7(a)**. The liquid fraction in the Laminar-CTO results continue to increase by 22.1% during the main iteration process, while it first increases by 32.3% followed by gradually decreasing by 12.8% in the Laminar-FGTO results, with the inflection point located at around 30 iteration steps, which is close to that in the upstream evolution process. Consequently, the final liquid fraction in the Laminar-CTO results is 5.1% higher in the upstream region, as shown in **Fig. 7(a)**, while is 6.1% lower in the downstream region, compared to the Laminar-FGTO results, as shown in **Fig. 7(b)**. Relative to the entire liquid cooling plate, the downstream region has higher heat dissipation demands and therefore inherently requires a greater allocation of solid phase. The upstream region inherently exhibits superior heat dissipation capabilities compared to the downstream region due to the inlet effect and lower coolant temperature, leading to a less requirement of solid phase. Therefore, such an optimization result by the Laminar-CTO, where it exhibits more solid fraction downstream and less solid phase upstream, demonstrates its enhanced regional selectivity and superior global planning capability, compared to the Laminar-FGTO.

As shown by the velocity and temperature gradient vectors in **Fig. 6**, optimal field synergy can be observed in the porous medium region where a clear solid-liquid interface has not yet formed. The optimization corresponds to such a process that, the uniform porous media is gradually separated into pure solid and liquid phases to reduce flow resistance, while the flow and temperature fields get distorted simultaneously, disrupting the nearly parallel velocity and temperature gradient in the original porous media, leading to a compromised field synergy and an increased synergy angle, as shown in **Fig. 6** and **Fig. 7(c)**. As squared in **Fig. 6(b)**, the rapidly established main flow channels in the Laminar-FGTO results are nearly straight, with velocity and temperature gradient vectors remain almost perpendicular that represents the worst field synergy. From this perspective, the competition of thermal-hydraulic objectives in TO can be reflected by the trade-off between field synergy and flow resistance among various structural topologies. A uniform porous media represents a better field synergy but worse hydraulic performance, while the straight flow channel represents a well-reduced flow resistance but compromised field synergy.

The consideration of field synergy by the CTO is conducive to more effectively balancing the competition of the thermal-hydraulic objectives, leading to an improved field synergy in the optimized liquid cooling plates. As demonstrated in **Fig. 7(c)**, during the optimization process where it inevitably involves sacrificing some field synergy in exchange for improved hydraulic performance, the incorporation of field synergy by the Laminar-CTO can effectively suppress the rate of increase in the synergy angle, i.e., balancing the competition between field synergy and flow resistance optimization. Consequently, during the iteration process, as shown in **Fig. 6**, the velocity and temperature gradient vectors in the Laminar-CTO results demonstrate better synergy, compared to the Laminar-FGTO results, with a 16% reduction in the average synergy angle of the optimized liquid cooling plate, as shown in **Fig. 7(c)**.

The Laminar-CTO and Laminar-FGTO achieve fragmented solid structure in the optimized liquid cooling plates, which contributes to enhance flow perturbation and disrupt continuous boundary layers, leading to an enhanced heat transfer performance. However, these fragmented structural topologies simultaneously leads to an expansion of flow stagnation zones as the inlet velocity increases, progressively compromising the enhancement on overall performance [26]. Consequently, the optimization results obtained in laminar flow regime with low *Re* struggle to sufficiently adapt to the high-*Re* turbulent flow conditions, and thus the Turbulent-CTO and Turbulent-FGTO optimized in turbulent flow regimes are also necessarily further discussed.

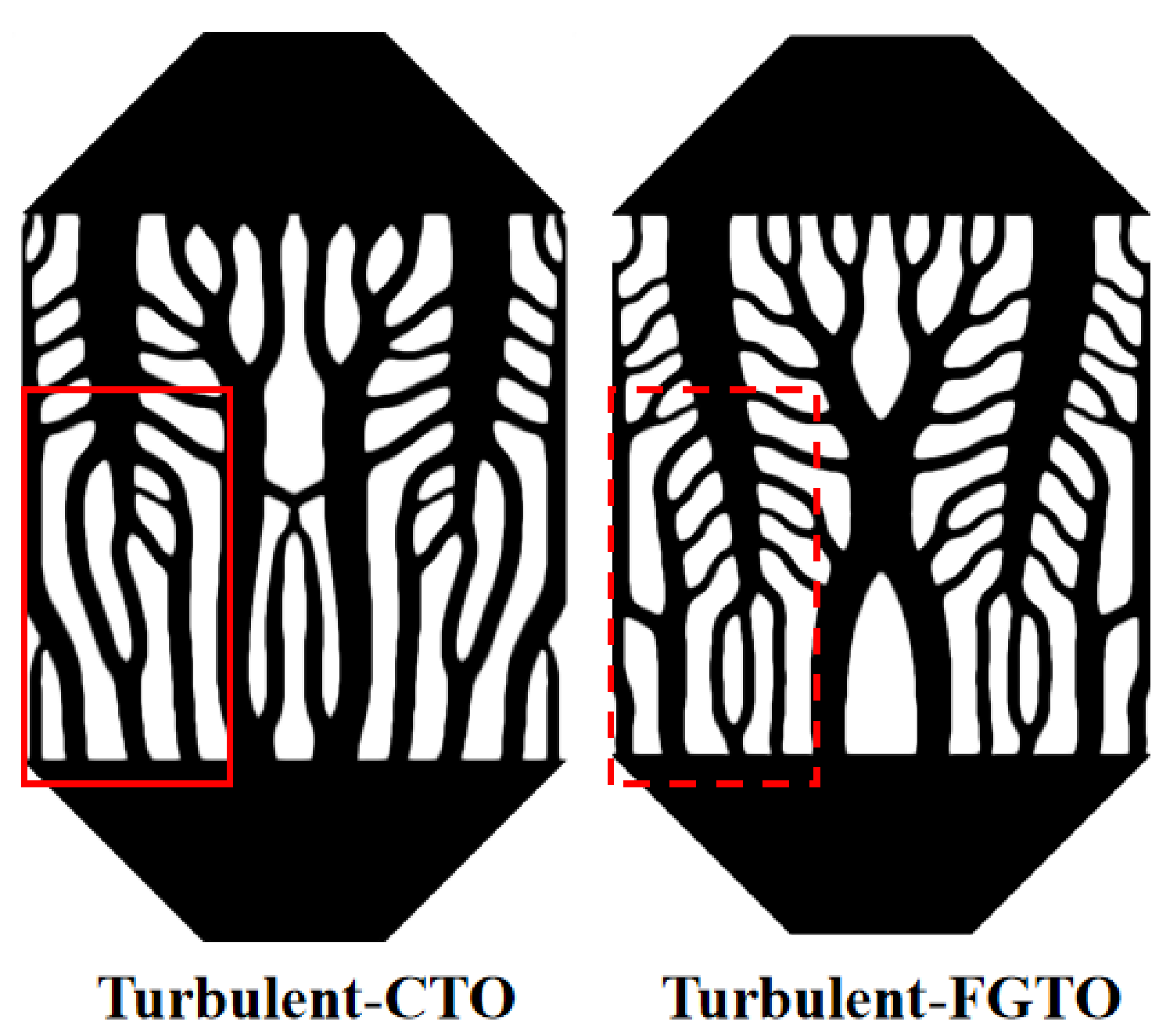

**Fig. 8.** The structural topologies of the optimized liquid cooling plates by the Turbulent-CTO and Turbulent-FGTO methods at $w_t$=0.7.

**Fig. 8** shows the structural topologies of the Turbulent-CTO and Turbulent-FGTO results at $w_t$=0.7. Compared with laminar TO, the effects of integrating field synergy on improving structural complexity are less pronounced for turbulent TO. For example, the heat transfer area of the Laminar-CTO result is 20.1% larger than that of the Laminar-FGTO result at $w_t$=0.7, while that of the Turbulent-CTO result even demonstrates a slight reduction of 9.6% compared to the Turbulent-FGTO result. It is further demonstrated in **Section 3.3** that the Turbulent-CTO focuses more on optimizing hydraulic performance while maintaining adequate thermal performance, different from the Laminar-CTO that enhances heat transfer performance at the cost of larger pressure drop. Consequently, although both Turbulent-CTO and Laminar-CTO can improve the PEC of the optimized liquid cooling plates, the improvement achieved by the Turbulent-CTO is less pronounced than that of the Laminar-CTO.

The diminished enhancement of incorporating field synergy into TO in turbulent flow regimes can be attributed to both the flow and heat transfer mechanism, and the inherent limitations of the field synergy theory and the solution to the turbulent TO problem as follows. The heat transfer in turbulent flow regime is dominated by the turbulence caused by eddies, which is distinct from the heat conduction through the thermal boundary layer under laminar flow conditions. Given that the improvement in field synergy achieved through structural topology primarily depends on generating secondary flows to enhance flow perturbations [63, 64], the benefits of structural optimization in improving synergy and heat transfer performance are minimal under turbulent flow conditions where flow perturbation are already sufficiently developed due to the strong turbulent mixing and eddy diffusion. Additionally, compared to the TO in laminar flow regime, the turbulent TO problem exhibits greater nonlinearity and computational instability, which elevates the difficulty to achieve high-performance structural topologies and makes it more prone to getting trapped in local optimal solutions [32]. Furthermore, except for the viscous sublayer region, the field synergy theory struggles to comprehensively characterize the convective heat transfer in turbulent flow due to its incapability to adequately account for the complex eddy effects [65], further constraining the superiority of the Turbulent-CTO over the Turbulent-FGTO.

**Table 5** Calculation time and iteration steps of the Laminar-CTO, Laminar-FGTO, Turbulent-CTO and Turbulent-FGTO at $w_t$=0.7.

| | Laminar-CTO | Laminar-FGTO | Turbulent-CTO | Turbulent-FGTO |
|---|---|---|---|---|
| Total time (hours) | 65.2 | 6.4 | 71.3 | 8.7 |
| Iteration steps | 227 | 163 | 192 | 181 |

It should be noted that since both the CTO and FGTO are conducted in the density-based TO framework, the inherent issues in the intermediate region (i.e., the gray area) still inevitably persist in the optimization results, despite the use of projection in Eq. (18). Nevertheless, as shown in **Fig. 5** and **Fig. S2**, except for a minor portion on the solid-liquid interface, these intermediate regions can be effectively eliminated in most area of the optimized liquid cooling plates. The average fractions of intermediate region across the entire liquid cooling plates only account for about 9.7% and 11.2% for the Laminar-CTO and Laminar-FGTO results at $w_t$=0.8, respectively. Consequently, the incorporation of field synergy theory by the CTO exhibits minimal influence on the gray area issue, and therefore it would not exert a significant impact on the optimization results nor the effectiveness of the CTO method.

Although incorporating field synergy theory into TO achieves a more rational structural topology distribution with the improved thermal-hydraulic performance, the highly coupled nature of the velocity, temperature and design variable fields significantly increases computational costs. As listed in **Table 5**, the Laminar-CTO and Turbulent-CTO require nearly one order of magnitude more computational time than the corresponding Laminar-FGTO and Turbulent-FGTO. Nevertheless, the introduction of field synergy theory demonstrates less deterioration on convergence, since there are no additional intermediate regions with poor convergence remaining in the optimization results, as shown in **Fig. 5**, and the total iteration steps does not exhibit a significant increase like the total computational time. Consequently, given the significant performance improvement by the incorporation of field synergy theory, the increased computational cost is acceptable.

### 3.2 Performance comparison under laminar flow conditions

The 3D numerical calculations are performed to evaluate the performance of the optimized liquid cooling plates. Performance of the CTO results are compared with that of the FGTO results to demonstrate and analysis the effect of incorporating field synergy into TO framework, and validating the superiority of the proposed CTO method over previous FGTO method. Additionally, the pin fin liquid cooling plate, as shown in **Fig. S3** of the **supplementary materials**, is adopted as a benchmark for performance comparison to provide an engineering reference for practical application.

**Fig. 9** shows the performance comparison results of the pin fin liquid cooling plate, the Laminar-CTO and Laminar-FGTO results at various $w_t$, at $V_{in}$=0.1 m/s, $T_{in}$=308.15 K and $q$=10 W/cm$^2$. For both the Laminar-CTO and Laminar-FGTO results, thermal performance is improved at the cost of larger pressure drop with increasing $w_t$, as shown in **Fig. 9(a)**, and the PEC is slightly increased with rising $w_t$ due to the more dominantly enhanced heat dissipation capabilities, as shown in **Fig. 9(b)**. Both the Laminar-FGTO and Laminar-CTO results demonstrates superior thermal and overall performances compared to the pin fin liquid cooling plate, while the pressure drops are larger due to the enhanced flow perturbation by complex structural topologies. The PEC of the Laminar-FGTO and Laminar-CTO results at $w_t$=0.7 are improved by 40% and 69%, respectively, compared to the pin fin liquid cooling plate, as shown in **Fig. 9(b)**, indicating a better overall thermal-hydraulic performance.

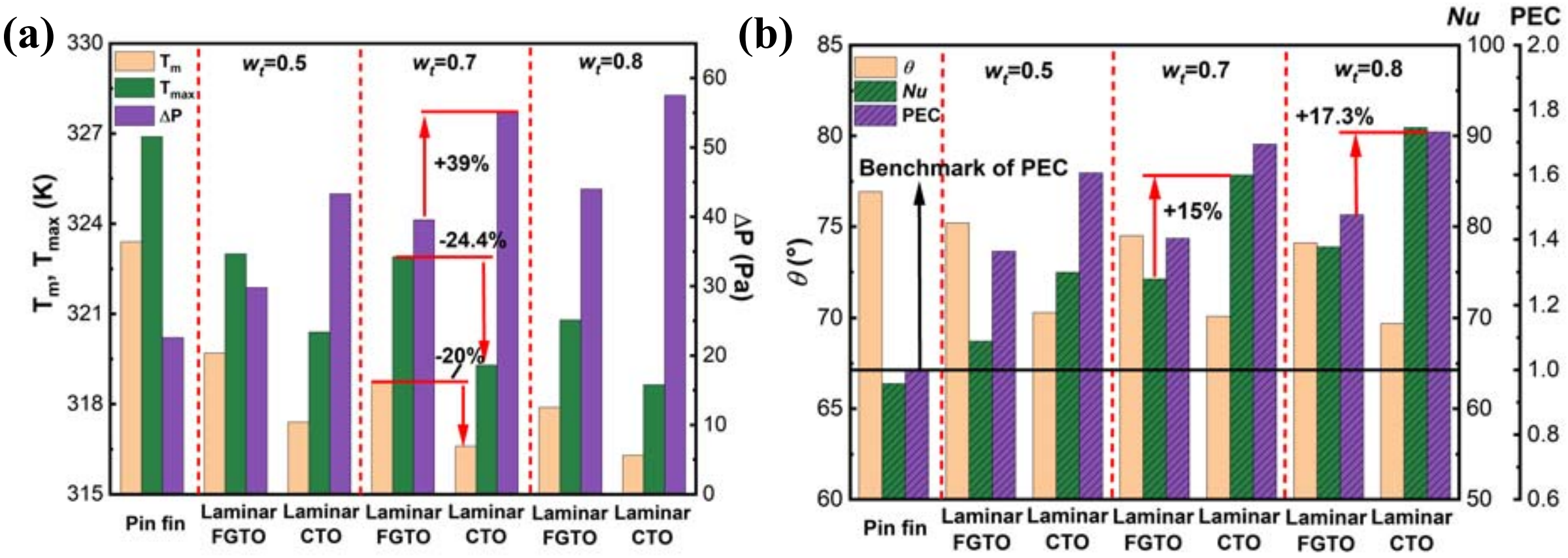

**Fig. 9.** Performance comparison results of various liquid cooling plates under the laminar flow condition of $V_{in}$=0.1 m/s, $T_{in}$=308.15 K and $q$=10 W/cm$^2$. (a) the average temperature of the heated surface $T_m$, the maximum temperature of the heated surface $T_{max}$, and the pressure drop $\Delta P$, (b) the synergy angle, Nusselt number and PEC

It can be observed from **Fig. 9(a)** that the Laminar-CTO can achieve the reduced average and maximum temperatures at various $w_t$, compared to the Laminar-FGTO, while the pressure drop is larger due to the enhanced flow perturbation arising from the more fragmented solid structures. For example, the average and maximum temperature rises of the Laminar-CTO result at $w_t$=0.7 are reduced by 20% and 24.4%, respectively, compared to the Laminar-FGTO result at $w_t$=0.7, while the pressure drop is increased by 39%. The hierarchical structural topologies in the Laminar-CTO result can enhance the synergy between flow and temperature fields, leading to an improved heat transfer performance. The average synergy angle of the Laminar-CTO result at $w_t$ =0.7 is reduced by 4.4°, and the $Nu$ is increased by 15%, compared with that of the Laminar-FGTO result at $w_t$ =0.7. Consequently, despite of higher flow resistance, the incorporation of field synergy theory can still improve the overall thermal-hydraulic performance, with the optimal PEC achieved by the Laminar-CTO being improved by up to 17.3% compared to the Laminar-FGTO at $w_t$ =0.8, as shown in **Fig. 9(b)**.

**Fig. 10(a)** and **Fig. 10(b)** show the temperature contours of the heated surface of the Laminar-CTO and Laminar-FGTO liquid cooling plates, respectively. It is demonstrated that the Laminar-CTO achieves improved thermal uniformity and reduced local high temperature, as marked in **Fig. 10**, compared to the Laminar-FGTO at various $w_t$. This can be attributed to the stronger interdependence of performance optimization across various regions achieved by incorporating overall field synergy into the objective function, which indirectly facilitates the uniform distribution of flow and heat dissipation capabilities across the entire liquid cooling plates.

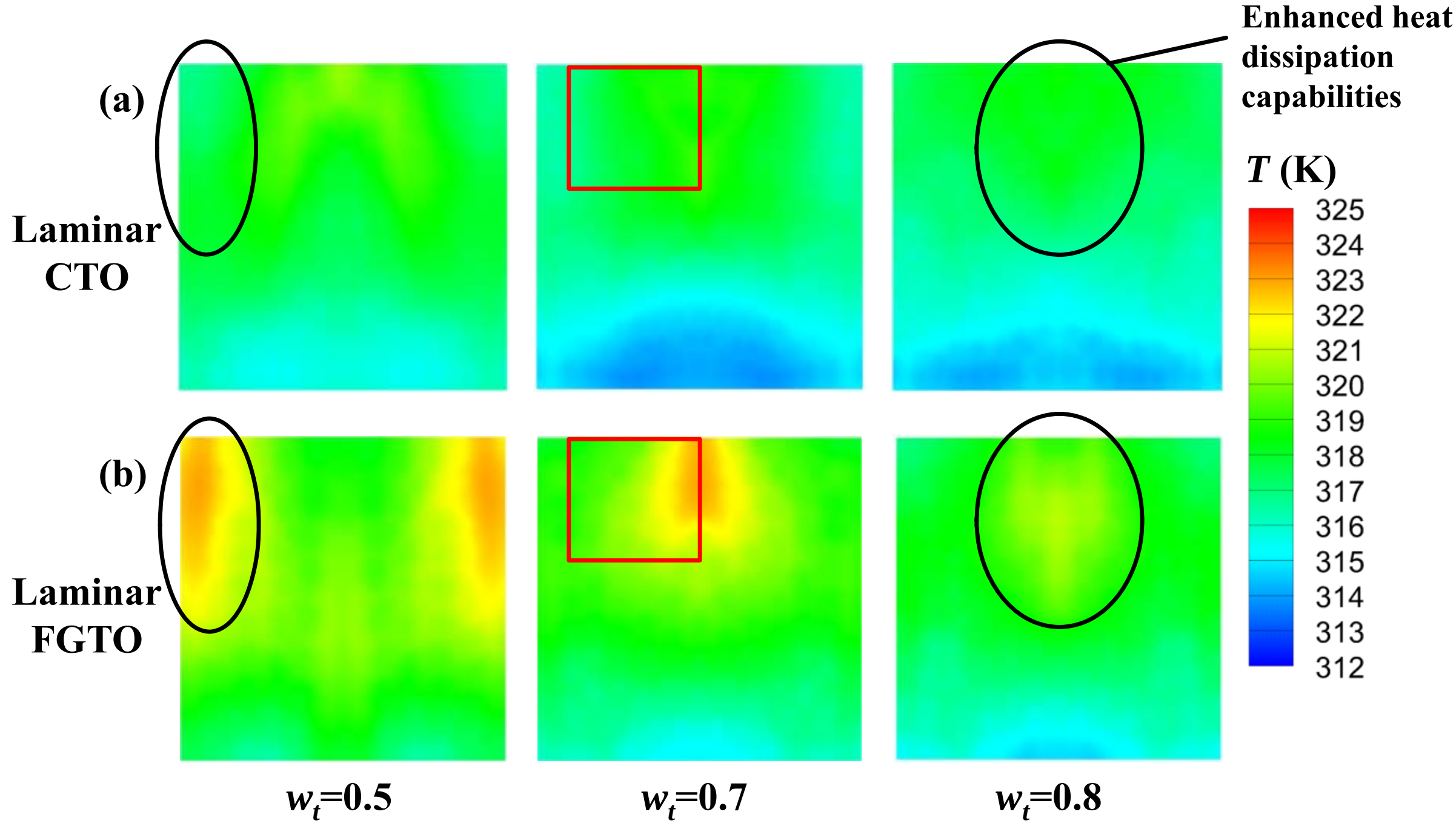

**Fig. 10.** The temperature contours of the heated surface of the Laminar-CTO and Laminar-FGTO liquid cooling plates at various $w_t$ at $V_{in}$=0.1 m/s, $T_{in}$=308.15 K and $q$=10 W/cm$^2$. (a) the Laminar-CTO results, (b) the Laminar-FGTO results

The mechanism by which the Laminar-CTO enhances thermal performance can be further understood by comparing the flow fields. It is well known that although fragmented structural topologies can effectively enhance flow perturbations to disrupt the boundary layer, they also introduce some low-velocity stagnation zones due to flow separation, which is not conducive to heat transfer but increases flow resistance [26]. This issue is particularly pronounced in TO liquid cooling plates, where complex structural topologies are densely arranged, and flow in branched channels might be compromised by these flow separation and stagnation zones induced by surrounding structures. Nevertheless, the incorporation of field synergy theory into TO can effectively mitigate this issue, since it naturally emphasizes on the synergistic enhancement of flow and heat transfer across various structural topologies rather than solely on local optimization, which is reflected in the optimized structural features that exhibit greater selective orientation and hierarchy.

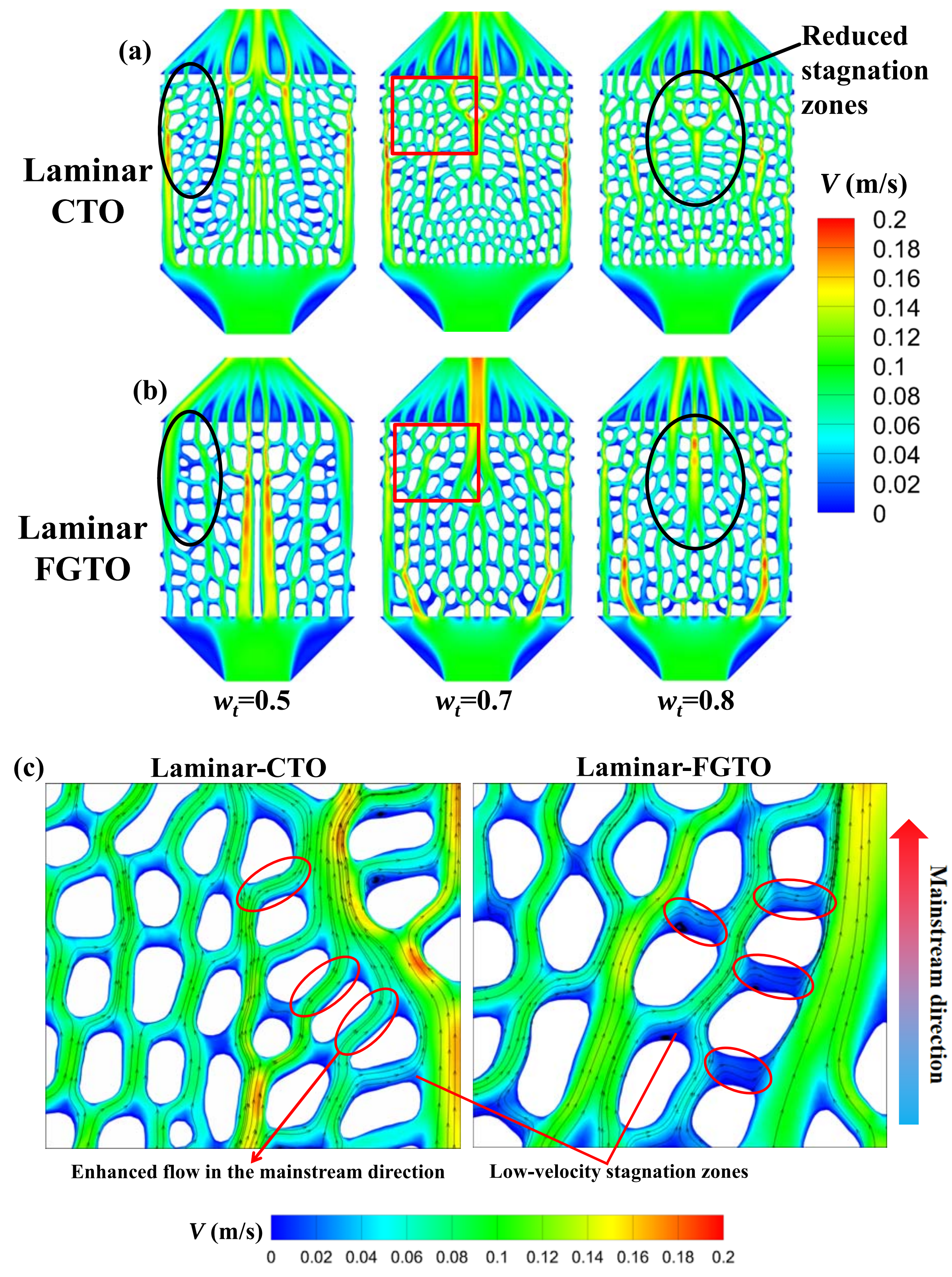

**Fig. 11.** Comparison of the velocity magnitude contours and streamlines on the middle planes in the direction of thickness of the Laminar-CTO and Laminar-FGTO liquid cooling plates at $V_{in}$=0.1 m/s. (a) the velocity magnitude contours of the Laminar-CTO results at various $w_t$, (b) the velocity magnitude contours of the Laminar-FGTO results at various $w_t$, (c) the local velocity magnitude contours and streamlines of the Laminar-CTO and Laminar-FGTO liquid cooling plates at $w_t$=0.7 which are the magnifications marked by the red squares in **Fig. 11(a)** and **Fig. 11(b)**

**Fig. 11(a)** and **Fig. 11(b)** show the velocity magnitude contours of the Laminar-CTO and Laminar-FGTO liquid cooling plates at various $w_t$, respectively. The high-temperature regions in the temperature contour shown in **Fig. 10** are marked at their corresponding locations in **Fig. 11** for a clearer demonstration. It is shown that the low-velocity stagnation zones in the Laminar-CTO results are significantly reduced, and flow in the branched channels is enhanced, compared to the Laminar-FGTO results, leading to the improved heat dissipation capabilities and reduced local high-temperature in the optimized liquid cooling plates, as shown in **Fig. 10**.

**Fig. 11(c)** shows the local velocity contours and streamlines of the Laminar-CTO and Laminar-FGTO results at $w_t$ =0.7 within the regions marked by red squares in **Fig. 11(a)** and **Fig. 11(b)**. It is demonstrated that the highly directional solids in the Laminar-CTO results not only conducive to reducing low-velocity stagnation zones, but also improve the direction of branched flow. In the Laminar-FGTO results without incorporation of field synergy theory, the flow in the branched channels is mostly nearly perpendicular to the main flow direction, and even exhibits a slight counter-flow trend, as shown by the red circles in **Fig. 11(c)**, which is detrimental to an efficient heat removal. Compared to the Laminar-FGTO results, the flow in the branched channels along the main flow direction is significantly enhanced in the Laminar-CTO results. Considering that the fluid is primarily heated along the main flow direction across the entire liquid cooling plate, the improved flow field by the Laminar-CTO can enhance synergy between temperature and velocity fields, leading to a better heat transfer performance.

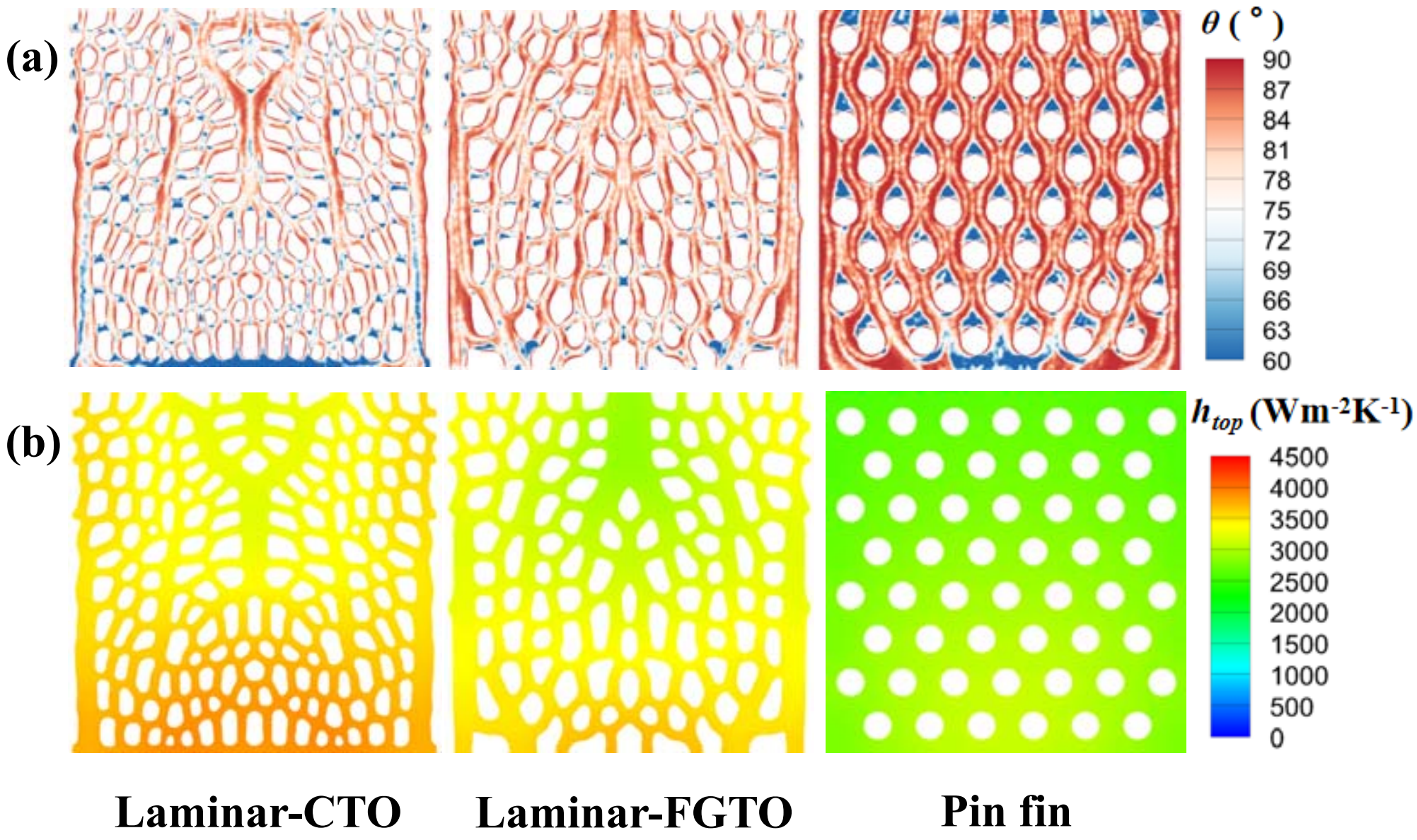

**Fig. 12.** The contours of the synergy angle and heat transfer coefficient of the Laminar-CTO and Laminar-

FGTO results at $w_t$ =0.7 and the pin fin liquid cooling plate at $V_{in}$=0.1 m/s, $T_{in}$=308.15 K and $q$=10 W/cm$^2$. (a) the synergy angle contours of the middle planes in the direction of thickness, (b) the heat transfer coefficient contours of the interface between the heated plate and fluid

**Fig. 12(a)** shows the synergy angle distribution of various liquid cooling plates on the middle planes in the direction of thickness. It can be observed that regions with high synergy angles tend to appear in coarse channels with fewer solid fins, while regions with low synergy angles tend to form in the flow separation zones. The synergy angle of the Laminar-FGTO results is more uniformly distributed than that in the pin fin liquid cooling plate, and the Laminar-CTO further improves the uniformity of the synergy angle distribution compared to the Laminar-FGTO, as shown in **Fig. 12(a)**. It has been proved that improving the distribution uniformity of synergy angle is conducive to reducing overall average synergy angle [63]. Consequently, as shown in **Fig. 9(b),** the synergy angle of the Laminar-FGTO results is reduced by 2.4° compared to the pin fin liquid cooling plate, while that of the Laminar-CTO results is further reduced by 4.4°, indicating a better synergy between flow and temperature fields.

**Fig. 12(b)** shows the heat transfer coefficient distribution of the pin fin liquid cooling plate, the Laminar-CTO and Laminar-FGTO results at $w_t$=0.7 on the interface between the heated plate and fluid, at $V_{in}$=0.1 m/s, $T_{in}$=308.15 K and $q$=10 W/cm$^2$. It can be observed that the heat transfer coefficients of various liquid cooling plates generally decrease from the inlet to outlet, due to the inlet effect and flow development. The complex structural topologies in the Laminar-FGTO results can significantly enhance heat dissipation capabilities compared to the pin fin liquid cooling plate. The improved field synergy by the hierarchical and directional structural topologies in the Laminar-CTO results further enhances the heat transfer performance in side and downstream regions, as shown in **Fig. 12(b)**, exhibiting superior overall heat transfer performance.

**Fig. 13** shows the performance comparisons of the pin fin liquid cooling plate, the Laminar-CTO and Laminar-FGTO results at $w_t$=0.7 at various inlet velocities. It is demonstrated that the incorporation of field synergy theory by the Laminar-CTO achieves the reduced average and maximum temperatures under various operating conditions compared to the pin fin liquid cooling plate and Laminar-FGTO results, as shown in **Fig. 13(a)** and **Fig. 13(b)**. Although the complex structural topologies lead to higher flow resistance and larger pressure drop compared to the pin fin liquid cooling plate, as shown in the **Fig. 13(c)**, both the Laminar-FGTO and Laminar-CTO methods can significantly improve the overall performance indicated by the PEC, as shown in **Fig. 13(d)**. However, it can be observed from **Fig. 13(d)** that this improvement is gradually compromised as the $V_{in}$ increases, with the PEC of the Laminar-FGTO and Laminar-CTO results decreased by 13.7% and 13.2%, respectively, as the $V_{in}$ increasing from 0.03 m/s to 0.2 m/s. This can be attributed to the optimization of the Laminar-FGTO and Laminar-CTO being conducted under laminar flow conditions, and the optimization fidelity gradually diminishes with increasing $V_{in}$. For example, the low-velocity flow stagnation zones, as shown in **Fig. 11**, would be augmented as the *Re* increases [26], leading to gradually diminishing enhancement in thermal performance and drastically rising pressure drop. Therefore, it is necessary to further evaluate performance of the Laminar-CTO and Laminar-FGTO results under turbulent flow conditions and investigate their differences compared to the Turbulent-CTO and Turbulent-FGTO, which are optimized in turbulent flow regimes.

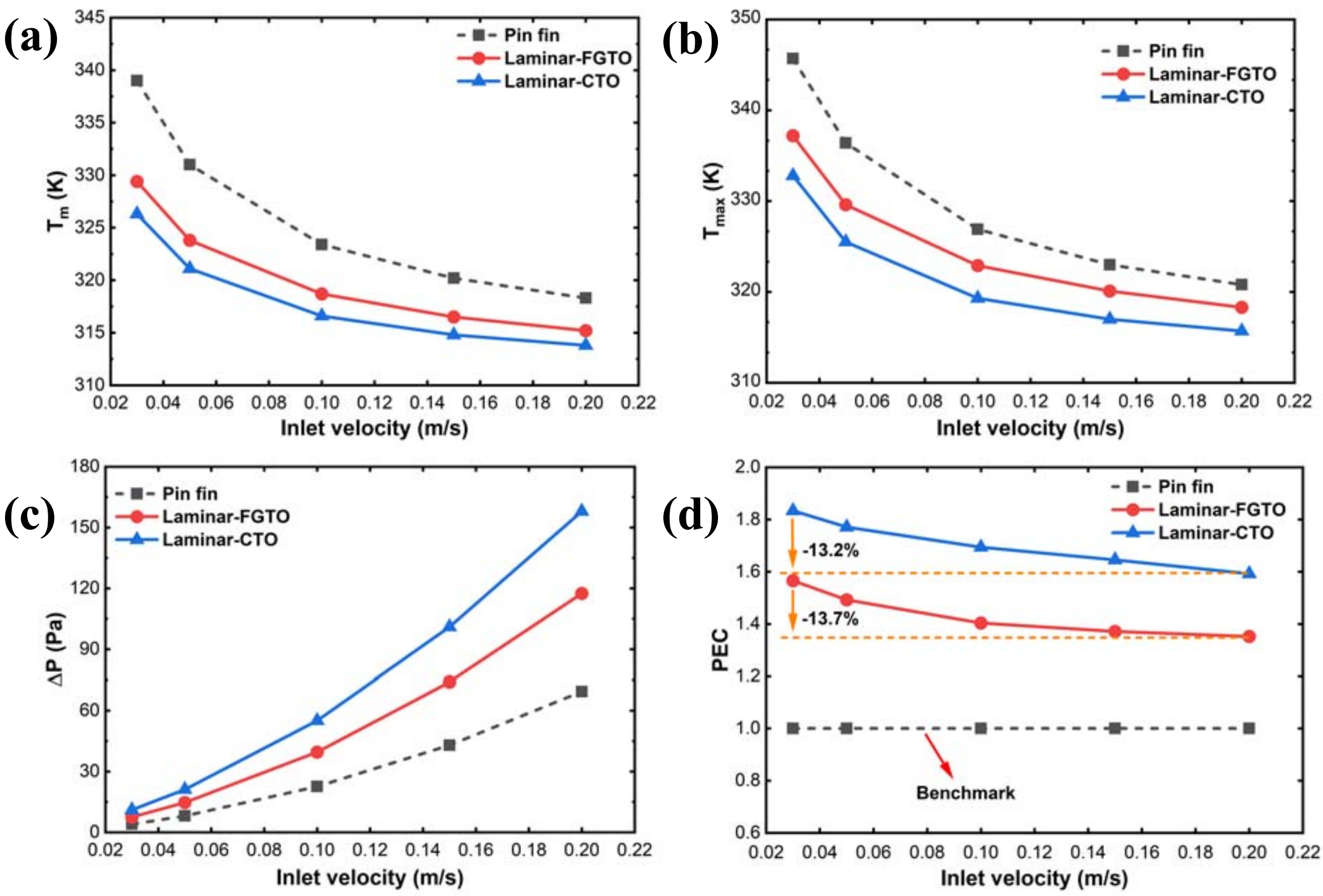

**Fig. 13.** The performance comparisons of the Laminar-CTO and Laminar-FGTO results at $w_t$=0.7 and the pin fin liquid cooling plate at various inlet velocities at $T_{in}$=308.15 K and $q$=10 W/cm$^2$. (a) the variation of $T_m$ with inlet velocity, (b) the variation of $T_{max}$ with inlet velocity, (c) the variation of $\Delta P$ with inlet velocity, (d) the variation of the PEC with inlet velocity

### 3.3 Performance comparison under turbulent flow conditions

In this section, thermal-hydraulic performance of the Laminar-CTO and Laminar-FGTO results optimized in laminar flow regime, and the Turbulent-CTO and Turbulent-FGTO results optimized in turbulent flow regimes, are all evaluated under turbulent flow conditions, with $V_{in}$ ranging from 0.7 m/s to 3.5 m/s (*Re* ranges from 5,500 to 38,500), $T_{in}$=308.15 K and $q$=100 W/cm$^2$. It is demonstrated that the Laminar-CTO and Turbulent-CTO demonstrate different features: the former significantly improves thermal performance but induces higher pressure drop, while the latter primarily enhances hydraulic performance with maintaining comparable thermal performance. Compared to Laminar-CTO, the further improvement in synergy and heat transfer performance in Turbulent-CTO is compromised. Nevertheless, both the Laminar-CTO and Turbulent-CTO can achieve superior overall thermal-hydraulic performances, i.e., the PECs, over those of the Laminar-FGTO and Turbulent-FGTO without incorporation of field synergy theory.

**Fig. 14** shows the performance comparisons of the pin fin liquid cooling plate, Laminar-FGTO, Laminar-CTO, Turbulent-FGTO and Turbulent-CTO results at $w_t$=0.7 under the turbulent flow conditions. **Fig. 15** shows the temperature and heat transfer coefficient distribution of various liquid cooling plates. It is demonstrated that all the TO liquid cooling plates can achieve superior thermal performance over the pin fin liquid cooling plate, with the improved temperature uniformity and heat transfer performance, as shown in **Fig. 15**. The average temperature rise of the Turbulent-CTO results is reduced by about 28% compared to the pin fin liquid cooling plate at $V_{in}$=1.4 m/s, as shown in **Fig. 14(a)**. Since the block-like fragmented structural topology in the Laminar-CTO results is more conducive to enhancing flow perturbation and improving heat transfer, thermal performance of the Laminar-CTO results is better than that of the Turbulent-CTO results, as shown in **Fig. 14(a)**, **Fig. 14(b)** and **Fig. 15**. Nevertheless, the strip-like structural topology in Turbulent-CTO exhibits superior capabilities in reducing flow resistance, and the pressure drop of the Turbulent-CTO results is reduced by up to 54% compared to the Laminar-CTO results, as shown in **Fig. 14(c)**. Furthermore, the strip-like solids generated in Turbulent-CTO results can effectively reduce the stagnation zones emerged at high-*Re* while improve the heat transfer area and reduce flow resistance [26]. Therefore, the Turbulent-CTO results are more applicable to high-velocity flow conditions, while the Laminar-CTO results are more favorable for low-velocity conditions [25, 32]. Consequently, the PEC of the Turbulent-CTO results is improved with rising $V_{in}$, while that of the Laminar-CTO result are gradually compromised as $V_{in}$ increases, as shown in **Fig. 14(d)**. It is demonstrated that the superiority of the Turbulent-CTO results becomes increasingly pronounced with increasing $V_{in}$, with the PEC of the Turbulent-CTO results being 10.5% lower than that of the Laminar-CTO results at $V_{in}$ = 0.7 m/s, while being 8.6% higher at $V_{in}$ = 3.5 m/s.

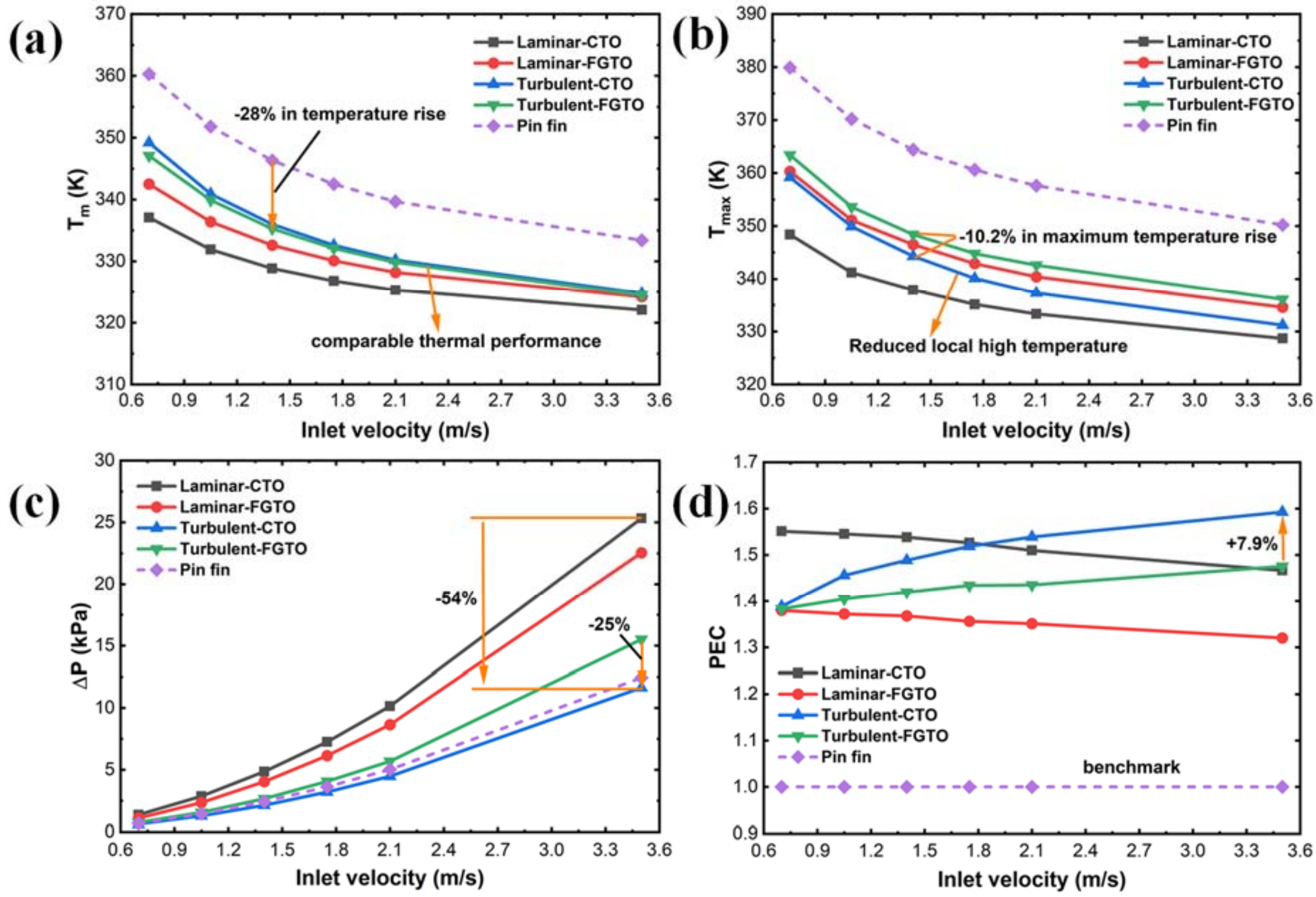

**Fig. 14.** The performance comparisons of the Turbulent-FGTO, Turbulent-CTO, Laminar-FGTO, Laminar-CTO results at $w_f$=0.7 and the pin fin liquid cooling plate under turbulent flow conditions at various inlet velocities at $T_{in}$=308.15 K and $q$=100 W/cm$^2$. (a) the variation of $T_m$ with $V_{in}$, (b) the variation of $T_{max}$ with $V_{in}$, (c) the variation of $\Delta P$ with $V_{in}$, (d) the variation with $V_{in}$

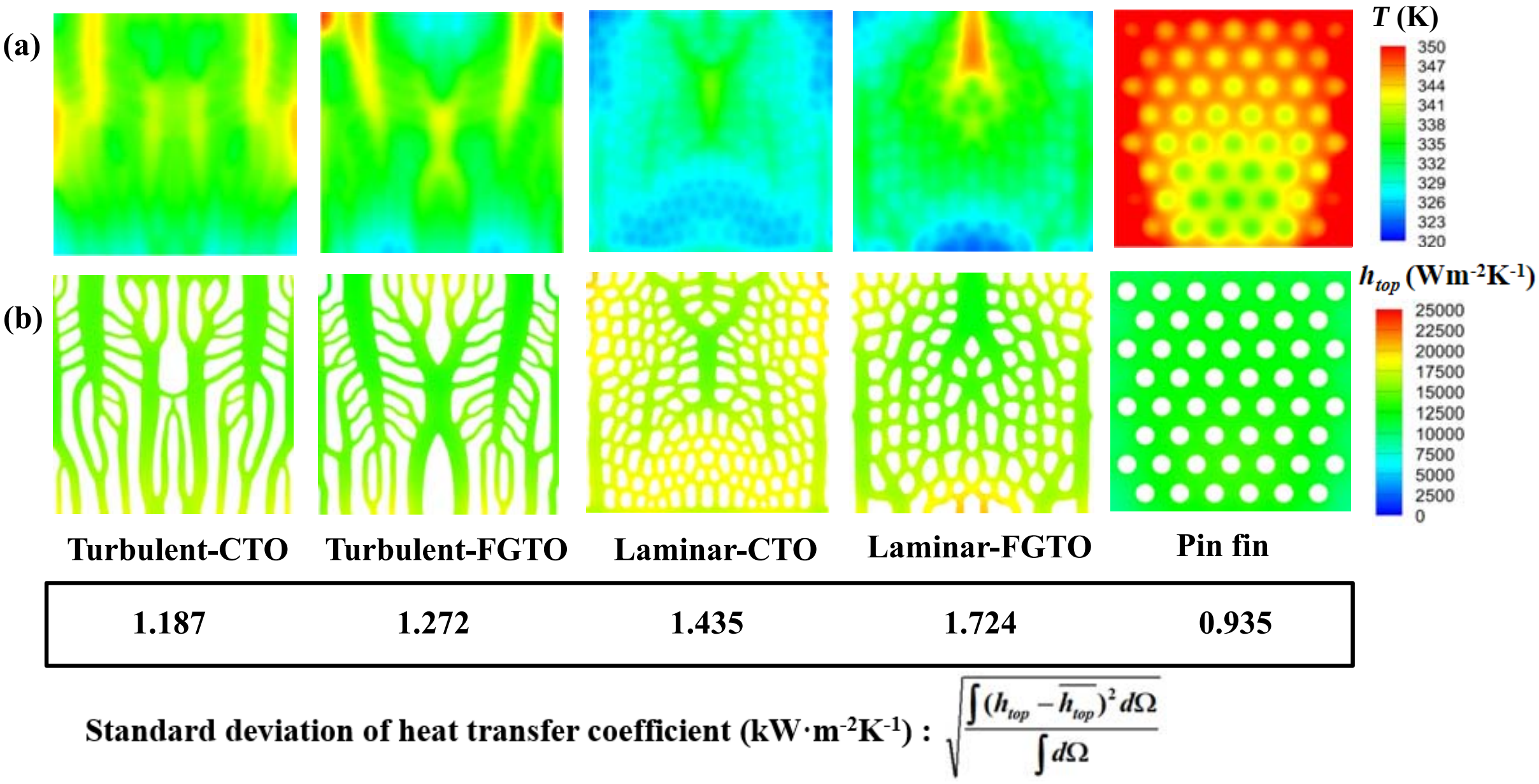

**Fig. 15.** The contours of the temperature and heat transfer coefficient of the Turbulent-CTO, Turbulent-FGTO, Laminar-CTO, Laminar-FGTO results at $w_f$=0.7 and the pin fin liquid cooling plate at $V_{in}$=1.4 m/s, $T_{in}$=308.15 K and $q$=100 W/cm$^2$. (a) the temperature contours of the heated surface, (b) the heat transfer coefficient

contours of the interface between the heated plate and fluid

Different from the Laminar-CTO, the average temperature of the Laminar-FGTO results can be significantly reduced by incorporating the field synergy, as discussed in **Section 3.2**, the Turbulent-CTO result exhibits comparable average temperature to the Turbulent-FGTO result. Nevertheless, the local hot spot in the Turbulent-CTO result is mitigated, as shown in **Fig. 15(a)**, with the maximum temperature rise reduced by 10.2% compared to the Turbulent-FGTO result at $V_{in}$=1.4 m/s, as shown in **Fig. 14(b).** Despite minimal improvement in thermal performance, the Turbulent-CTO can reduce the pressure drop by up to 25% compared to the Turbulent-FGTO, as shown in **Fig. 14(c)**, indicating the Turbulent-CTO focuses more on improving hydraulic performance while maintaining comparable thermal performance, which is distinct from the Laminar-CTO that improves thermal performance at the cost of larger pressure drop. Consequently, the overall thermal-hydraulic performance, i.e., the PEC, of the Turbulent-CTO result can be improved by 7.9% compared to the Turbulent-FGTO result at $V_{in}$=3.5 m/s, as shown in **Fig. 14(d)**.

**Fig. 16(a)** and **Fig. 16(b)** show the variations of average heat transfer coefficient and average synergy angle with $V_{in}$, respectively, of various liquid cooling plates. **Fig. 16(c)** and **Fig. 16(d)** show the distributions of synergy angle and velocity magnitude of the Turbulent-CTO and Turbulent-FGTO results, respectively. It is demonstrated that the Turbulent-CTO only exhibits slight improvement in the field synergy, as shown in **Fig. 16(c)**, thus on the heat transfer performance. The average heat transfer coefficient of the Turbulent-CTO result is improved by 5.9% compared to the Turbulent-FGTO, while the Laminar-CTO achieves an 11.4% improvement compared to the Laminar-FGTO, at $V_{in}$=1.4 m/s, as shown in **Fig. 16(a)**. The synergy angle of the Turbulent-CTO result is only 1.3° lower than that of the Turbulent-FGTO results, while the Laminar-CTO can reduce it by 4.2° compared to the Laminar-FGTO, as shown in **Fig. 16(b)**. These further prove that the optimization of heat transfer performance is compromised in turbulent flow regimes. As shown in **Fig. 15(b)**, the heat transfer coefficient in the Turbulent-CTO and Turbulent-FGTO results are more uniformly distributed compared to the Laminar-CTO and Laminar-FGTO, e.g., the standard deviation of heat transfer coefficient of the Turbulent-FGTO is 26.2% lower than that of the Laminar-FGTO. Additionally, both the Turbulent-CTO and Laminar-CTO tend to improve the distribution uniformity of heat transfer coefficient, compared to the Turbulent-FGTO and Laminar-FGTO without incorporation of field synergy theory, respectively. Consequently, it is indicated that for the Turbulent-CTO and Turbulent-FGTO that inherently tend to generate structural topologies with high heat transfer uniformity, additional incorporation of field synergy theory might impose less optimization potential for further improvement of synergy and heat transfer performance, compared to the Laminar-CTO and Laminar-FGTO. Although the structural complexity of the Turbulent-CTO is slightly inferior to that of the Turbulent-FGTO, as shown in **Fig. 8**, the Turbulent-CTO can still effectively enhance flow with higher velocity in the branched channels, as shown in **Fig. 16(d)**, leading to a higher heat transfer coefficient, as shown in **Fig. 16(a)**. Consequently, despite a 9.6% reduction of heat transfer area, the Turbulent-CTO can still achieve comparable thermal performance while significantly improving hydraulic performance, compared to the Turbulent-FGTO, leading to an improved PEC, as shown in **Fig. 14(d)**.

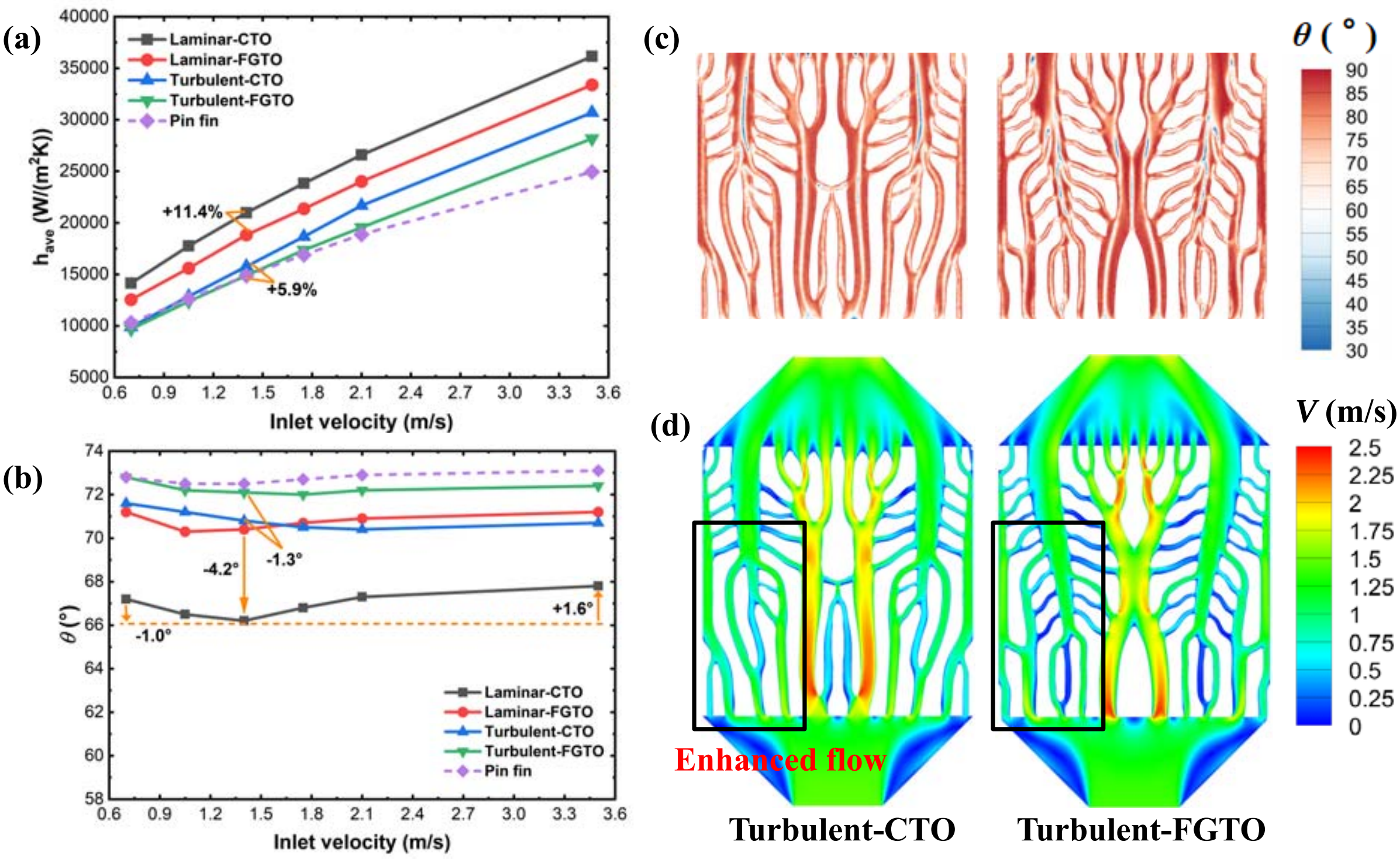

**Fig. 16.** Enhancement by the Turbulent-CTO compared to the Turbulent-FGTO on flow, synergy and heat transfer performance at $T_{in}$=308.15 K and $q$=100 W/cm$^2$. (a) the variation of average heat transfer coefficient $h_{ave}$ of various liquid cooling plates with inlet velocity, (b) the variation of synergy angle of various liquid cooling plates with inlet velocity, (c) synergy angle distribution on the middle plane of the Turbulent-CTO and Turbulent-FGTO results at $V_{in}$=1.4 m/s, (d) velocity magnitude contours on the middle plane of the Turbulent-CTO and Turbulent-FGTO results at $V_{in}$=1.4 m/s

The average synergy angles of various liquid cooling plates all exhibit an initial reduction with increasing $V_{in}$, and then followed by a gradual leveling off with a slight rise, as shown in **Fig. S4** and **Fig. 16(b)**. The synergy angle of the Laminar-CTO result is reduced by 3.1° as the $V_{in}$ increases from 0.03 m/s to 0.2 m/s, as shown in **Fig. S4**, while only reduced by 1.0° as the $V_{in}$ increases from 0.7 m/s to 1.4 m/s, as shown in **Fig. 16(b)**, and it slightly rises by 1.6° as $V_{in}$ increases from 1.4 m/s to 3.5 m/s. Under low-velocity condition, an increase in $V_{in}$ contributes to improve the synergy between the velocity and temperature fields [64]. Under high-velocity condition where the perturbation effects caused by the complex structural topologies are sufficiently developed, further increasing $V_{in}$ provides minimal improvement in synergy and might even partially disrupt it due to the expansion of flow stagnation zones [35, 64]. It is further suggested that compared to the Laminar-CTO, the Turbulent-CTO has less potential for further improving synergy and heat transfer performance due to the adequately enhanced flow and stabilized synergy in turbulent flow regime.

### 3.4 Manufacturability verification of the CTO liquid cooling plate

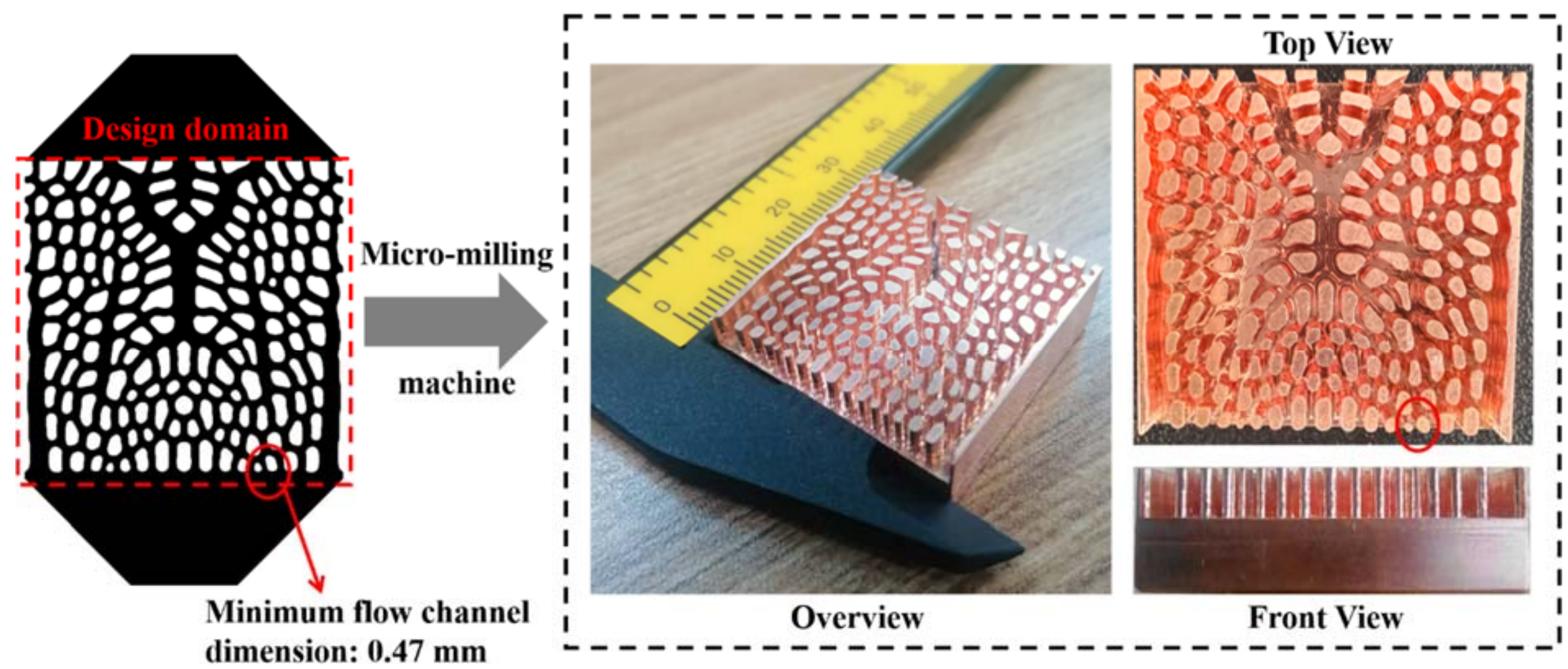

**Fig. 17.** Photographs of the fabricated prototype of the Laminar-CTO liquid cooling plate at $w_t$=0.7

Although the accuracy of the 3D numerical calculation model has been verified in **Section 2.8**, manufacturing uncertainties would inevitably introduce discrepancies between the ideal optimized structure used in the numerical calculations and the fabricated structure for practical application. Given the high complexity of the CTO liquid cooling plate, further manufacturability tests are essential to verify the feasibility of the proposed CTO method. The Laminar-CTO liquid cooling plate at $w_t$=0.7 with the highest structural complexity and manufacturing cost is selected as a representative for the manufacturability verification. **Fig. 17** shows the photographs of the fabricated liquid cooling plate. It is demonstrated that the CTO liquid cooling plate can be readily machined through the computer numerical control micro-milling with a dimensional tolerance of ±0.02 mm. With a minimum flow channel dimension of 0.47 mm, as circled in **Fig. 17**, dimensional uncertainty introduced during manufacturing can be controlled within ±4.2%. Consequently, the dimensional deviation exerts minimal impact on the achieved performance, further demonstrating the feasibility and effectiveness of the CTO method.

This study focuses on the theory and methodology, integrating the field synergy and fractal geometry theories to construct the CTO framework that reflects comprehensive mechanism of convective heat transfer in optimization. The effectiveness of the proposed CTO method and the manufacturability of the optimized liquid cooling plate have been verified as a first-step work, and the systematic experimental investigation will be the focus of the next-step work. Furthermore, the CTO method proposed in this study is currently limited to the optimization of liquid cooling plates, and the future research topics include the optimization of a variety of heat transfer devices, e.g., fin designs for energy storage devices by the CTO.

## 4. Conclusion

In this study, a convective heat transfer topology optimization (CTO) method is proposed, where the field synergy theory and fractal geometry theory are integrated into the density-based TO framework to explicitly depict the heat transfer coefficient and heat transfer area in the thermal objective function, achieving a direct optimization of convective heat transfer. The CTO is compared with the FGTO method without incorporating field synergy theory, under both laminar and turbulent flow conditions. The main conclusions are as follows:

(1) The CTO tends to generate more hierarchical and directional structural topologies, which contributes to reduce flow stagnation zones and enhance heat transfer performance through improving the flow direction in the branched channels and field synergy. Under laminar flow conditions, the Laminar-CTO can improve the *Nu* by 15% while reducing the average temperature rise by 20%, compared to the Laminar-FGTO. Under turbulent flow conditions, the Turbulent-CTO can reduce pressure drop by 25% while reducing the maximum temperature rise by 10.2%, compared to the Turbulent-FGTO.

(2) The incorporation of field synergy theory into laminar and turbulent TOs demonstrates different impacts on optimization results: The Laminar-CTO can significantly improve the synergy and heat transfer performance but lead to higher flow resistance; the Turbulent-CTO focuses more on improving hydraulic performance while maintaining comparable thermal performance. Compared with the Laminar-CTO, the Turbulent-CTO exhibits compromised enhancement on synergy and heat dissipation capabilities, since the synergy tends to stabilize under the turbulent flow conditions with sufficiently enhanced flow perturbations. Nevertheless, both the Laminar-CTO and Turbulent-CTO can achieve superior overall thermal-hydraulic performance, with the PEC of the Laminar-CTO results improved by 17.3%, and that of the Turbulent-CTO results improved by 7.9%, compared to the Laminar-FGTO and Turbulent-FGTO results, respectively.

(3) The improvement by the CTO is directly reflected in the fact that the incorporation of field synergy into optimization contributes to spontaneously reduce the optimized structures that simply stack solids locally without favoring overall heat transfer, thereby achieving proactive avoidance of local optimal solutions in early stage of iteration. Despite a significant performance improvement, the CTO suffers from the heavy computational costs due to the highly coupled design variable, flow and temperature fields in the objective function.

**CRediT authorship contribution statement**

Zixu Han: Writing – original draft, Methodology, Investigation, Formal analysis. Peng Zhang: Writing – review & editing, Validation, Supervision, Funding acquisition, Conceptualization.

**Declaration of competing interest**

The authors declare that they have no known competing financial interests or personal relationships that could have appeared to influence the work reported in this paper.

**Data availability**

Data will be made available on request

**Acknowledgements**

This research is supported by the Natural Science Foundation of Shanghai Municipality under the Contract No. 25ZR1401209. This research is also partially supported by the National Natural Science Foundation of China under the Contract No. 52576220. A few characterizations are conducted at the AEMD of Shanghai Jiao Tong University.